\documentclass[prb,aps,twocolumn, amssymb,amsmath,floatfix,superscriptaddress,nofootinbib]{revtex4-1}

\bibpunct{[}{]}{;}{n}{}{}

\usepackage[utf8]{inputenc}
\usepackage{tabularx}
\usepackage{bm}
\usepackage{euscript}
\usepackage{epsfig,psfrag}
\usepackage{graphicx}
\usepackage{color}
\usepackage{amsfonts}
\usepackage{exscale}
\usepackage{amsbsy}
\usepackage{wrapfig}

\usepackage{amsmath}
\usepackage{enumitem}
\usepackage{natbib}
\usepackage{verbatim}
\setcitestyle{square}

\usepackage{hyperref}
\hypersetup{
	colorlinks=true,       
    linkcolor=red,          
    citecolor=blue,        
    filecolor=magenta,      
    urlcolor=black,           
    pdfauthor={Author},     
}

\usepackage{slashed} 
\numberwithin{equation}{section} 
\renewcommand{\theequation}{\arabic{section}.\arabic{equation}} 

\newcommand{\bea}{\begin{eqnarray}}  
\newcommand{\eea}{\end{eqnarray}}
\newcommand{\ben}{\begin{enumerate}}
\newcommand{\een}{\end{enumerate}}
\newcommand{\be}{\begin{equation}}
\newcommand{\ee}{\end{equation}}

\renewcommand{\theequation}{\arabic{section}.\arabic{equation}}

\begin{document}
\title{Operator Growth and Symmetry-Resolved Coefficient Entropy \\ in Quantum Maps}

\author{Laimei Nie}
\affiliation{Department of Physics and Institute for Condensed Matter Theory, University of Illinois at Urbana-Champaign, Urbana, Illinois 61801, USA}

\begin{abstract}

Operator growth, or operator spreading, describes the process where a ``simple" operator acquires increasing complexity under the Heisenberg time evolution of a chaotic dynamics, therefore has been a key concept in the study of quantum chaos in both single-particle and many-body systems. An explicit way to quantify the complexity of an operator is the Shannon entropy of its operator coefficients over a chosen set of operator basis, dubbed ``coefficient entropy"\cite{Shivaji2019}. 
However, it remains unclear if the basis-dependency of the coefficient entropy may result in a false diagnosis of operator growth, or the lack thereof.  
In this paper, we examine the validity of coefficient entropy in the presence of hidden symmetries. Using the quantum cat map as an example, we show that under a generic choice of operator basis, the coefficient entropy proposed in \cite{Shivaji2019} fails to capture the suppression of operator growth caused by the symmetries. We further propose ``symmetry-resolved coefficient entropy" as the proper diagnosis of operator complexity, which takes into account robust unknown symmetries, and demonstrate its effectiveness in the case of quantum cat map. 
\end{abstract}

\maketitle

\section{Introduction}
Many-body quantum chaos has captured extensive attention from condensed matter, high energy, and quantum information communities due to its potential connections with quantum thermalization ~\cite{Deutsch1991, Srednicki1994}, many-body localization~\cite{Basko2006MBL, PalHuse2010MBL}, and black hole physics~\cite{RobertsStanford2015diagnosing, ShenkerStanford2014butterfly, Maldacena2016bound, Cotler2017RMT}.  Among the many operational diagnostics of quantum chaos, operator growth, or operator spreading, has played a vital role in characterizing dynamical chaos in the operator Hilbert space. It describes the process where a ``simple" operator becomes increasingly complicated under chaotic Heisenberg time evolution, and can be intuitively quantified by out-of-time-ordered correlator (OTOC) \cite{Larkin1969, Maldacena2016bound}, operator entanglement \cite{Dubail2017OpEE, Chen2018, Kudler-Flam2021local}, Krylov complexity \cite{Parker2019Krylov, Kar2021KrylovGravity, Caputa2021KrylovCFT}, and coefficient entropy \cite{Shivaji2019}. 
The common belief, which has been tested in various discrete and continuous models, is that for a quantum dynamics to be chaotic it must satisfy certain form of maximal operator growth.

The primary focus of this paper will be on the coefficient entropy, first explicitly proposed and evaluated in the context of single particle quantum mechanical models but also has many-body counterpart\cite{Roberts2018growth, Qi2019Measurement}. 
Generally speaking, the coefficient entropy of a time-evolving operator is defined as the Shannon entropy of its expansion coefficients over a chosen set of operator basis. 
For an initially ``simple" operator consisting of superposition of one or few basis operators, under a chaotic dynamics its coefficient entropy is supposed to grow to its largest possible value set by the Hilbert space dimension.
However, such definition depends on the choice of basis, and it remains a question if such ambiguity jeopardizes the validity of coefficient entropy. Previous study \cite{Shivaji2019} on certain single-particle quantum mechanical models showed that the long-time value of the coefficient entropy is not affected by basis transformation, leading to the conjecture that quantum chaos is characterized by the long-time saturation of coefficient entropy for most choices of operator bases. However, here we will show in a similar but more generic set of models that the coefficient entropy can be highly susceptible to the choice of basis operators, thus cannot serve as a proper diagnostics of operator growth and quantum chaos.

Specifically, the coefficient entropy fails when there are {\it unknown} symmetries, a common cause for the lack of maximal operator growth. In this case the Hamiltonian possesses a block-diagonal structure, and the operator growth is limited to each symmetry block. Without the knowledge of the symmetries, one would wrongly conclude, from the coefficient entropy measured under a generic choice of operator basis, that the dynamics supports maximal operator growth. To overcome this, we propose  ``symmetry-resolved coefficient entropy" as the unambiguous diagnostics. We provide a recipe that block-diagonalizes the Hamiltonian with hidden symmetries, and evaluate the coefficient entropy within each symmetry subspace. We demonstrate the effectiveness of our method in the quantum cat map, a well-studied single-particle quantum mechanical model that can be tuned between quantum chaotic and non-chaotic regimes via the inclusion of non-linear perturbations and quantum symmetries. The symmetry-resolved coefficient entropy in such systems precisely captures the existence of the symmetries and agrees with the prediction from level statistics revealing sub-maximal operator growth.

We would also like to briefly comment on the distinction and connection between quantum chaos in single-particle and many-body systems. It is commonly accepted that interactions play a crucial role in generating a complicated, correlated energy spectrum and in consequence, thermalization and chaos in many-body quantum systems. Nevertheless, some of the diagnostics of many-body quantum chaos are inspired by, or tightly linked to, the chaos diagnostics in single-particle systems
which lack interactions or tensor-product structure in Hilbert space. For instance, the random matrix theory aspect of many-body quantum chaos ~\cite{Mondaini2016IsingRMT, Cotler2017RMT, Balasubramanian2017echos, Cotler2017complexity, Prosen2018RMT, Benjamin2019irrational} is motivated by the Bohigas-Giannoni-Schmit conjecture~\cite{Bohigas1984} which was originally proposed and tested in single-particle models, and the study of many-body quantum scars~\cite{Turner2018scar, Moudgalya2018scar, Lin2019scar} is partially inspired by the scarring behavior of eigenfunctions in non-interacting systems. Through the study of quantum cat maps, we hope to highlight the 
importance of hidden symmetries in the characterization of operator growth, and shed light on the understanding of operator size and quantum chaos in many-body systems.

The paper is organized as follows. In Section~\ref{sec: GeneralConsiderations}, we review the definition of coefficient entropy and point out how its dependency on the choice of operator basis may lead to a false diagnosis of operator growth in the presence of symmetries. 
We then introduce as the proper diagnostics the symmetry-resolved coefficient entropy, and provide a recipe that enables one to block-diagonalize a generic Hamiltonian in the presence of robust unknown symmetries.
Next, we test our framework in the prototypical single-particle model, the quantum cat map, introduced in Section~\ref{sec: QuantumCatMaps}. 
We start by reviewing the classical Arnold's cat map and its perturbed version, and introduce the corresponding quantum cat maps based on Hannay and Berry's quantization scheme~\cite{Hannay-Berry}. We then present a particular set of quantum cat maps with additional symmetries that will be used to demonstrate the limitation of coefficient entropy and more importantly, the effectiveness of the symmetry-resolved coefficient entropy.
Section~\ref{sec: CEinQCatMap} contains the result of coefficient entropy in such quantum cat maps. We show, both numerically and analytically, that under a generic choice of operator basis, the long-time saturation value of the coefficient entropy fails to capture the existence of the symmetries, which manifest themselves in the Poisson-like level statistics. 
In Section~\ref{sec: SRCEinQCatMap}, following the generic procedure described in Section~\ref{sec: GeneralConsiderations} we evaluate the symmetry-resolved coefficient entropy in the same system. Its long-time saturation value in a particular symmetry block is shown to be independent of the choice of basis and reveals a maximal operator growth, in agreement with the Wigner-Dyson level statistics. 
Finally, in Section~\ref{sec: conclusion} we discuss the implication of our framework in other diagnostics of operator growth, as well as in the context of many-body quantum chaos, and raise questions for future investigations.

\section{General Considerations}
\label{sec: GeneralConsiderations}
\subsection{Review of Coefficient Entropy}
The coefficient entropy was introduced in~\cite{Shivaji2019} as a means to quantify the complexity of an operator. Given a $D$-dimensional operator Hilbert space equipped with a unitary quantum dynamics, an initial operator $\mathcal{O}$ under Heisenberg time evolution can be expanded using a set of orthonormal operator basis $\{ \mathcal{B}_n \}$
\be
\mathcal{O}(t) = \sum\limits_{n=1}^D c_n(t) \mathcal{B}_n,
\ee
where $\{ \mathcal{B}_n \}$ satisfy $\mbox{Tr} (\mathcal{B}^{\dagger}_m \mathcal{B}_n) = D \delta_{mn}$. The coefficients $\{c_n(t) \}$, normalized via $\sum\limits_{n} |c_n(t)|^2 = 1$, contain the information of the ``weight" distribution of $\mathcal{O}(t)$ over the basis operators.
More precisely, one can define the coefficient entropy to quantify how much $\mathcal{O}(t)$ has spread in the operator Hilbert space under the time evolution:
\be
S_{\mathcal{O}}(t) \equiv -\sum\limits_{n=1}^{D} |c_n|^2 \log |c_n|^2.
\label{eq: CoeffEntropy}
\ee
It can be easily seen that $S_{\mathcal{O}}(t)$ satisfies the following inequality
\be
0 \le S_{\mathcal{O}}(t) \le \log D,
\ee
where the lower bound is a consequence of the non-negativity of entropy, and the upper bound is reached when $|c_n|^2 = 1/D$ for all $n$'s. For a fixed choice of basis, a ``simple" operator is characterized by a zero or small coefficient entropy, whereas a maximally complicated operator is defined by its coefficient entropy reaching the largest possible value. Roughly speaking, a chaotic dynamics should be able to bring simple operators to complex ones over the course of time.

While the coefficient entropy brings us an intuitive picture of understanding operator growth, one caveat associated with its definition is the basis-dependency. Different choices of basis operators may lead to different long-time values of the coefficient entropy, and it remains unclear which one should serve as the diagnostics.
(Note that there is a special set of operator basis under which the coefficient entropy remains constant over time: the operators $\{ |\phi_m \rangle \langle \phi_n | \}$ constructed from the eigenstates $\{ |\phi_m \rangle \}$ of the dynamics. We exclude them from our choice of operator basis.)
It was conjectured in~\cite{Shivaji2019} that for a quantum chaotic dynamics, the coefficient entropy of a simple operator should saturate the maximal value for most choices of operator basis. This conjecture was tested in a family of quantum cat maps, where despite a drastic change in the early-time behavior upon a random basis transformation, the coefficient entropies saturate to the same value at late time.

However, as we shall see, even within the scope of quantum cat maps there are counterexamples where the coefficient entropy is not capable of identifying the suppression of full operator growth. Specifically, these quantum cat maps possess additional symmetries that have not been previously considered in~\cite{Shivaji2019}. On a more general consideration, the common existence of hidden symmetries in quantum dynamics has important impact on the energy spectrum hence the level of chaos, therefore one needs a framework that takes care of those symmetries before proceeding to the evaluation of operator growth.

\subsection{Symmetry-Resolved Coefficient Entropy}
\label{subsec: SRCE}
To properly characterize operator growth and quantum chaos in the presence of symmetries, we introduce the symmetry-resolved coefficient entropy. Suppose the time evolution operator has a hidden block-diagonal structure due to an unknown symmetry. Within one such block (symmetry subspace) of dimension $\tilde D$, 
an operator $\tilde{\mathcal{O}} (t)$ can be expanded using a set of orthonormal operator basis $\{  \tilde{\mathcal{B}}_m \}$
\be
\tilde{\mathcal{O}} (t) = \sum\limits_{m = 1}^{\tilde D} \tilde{c}_m(t) \tilde{\mathcal{B}}_m.
\ee
Note that at the moment this is solely a formal expansion as we do not know the explicit form of the symmetry.
The symmetry-resolved coefficient entropy of $\tilde{\mathcal{O}}$ is then defined as
\be
S^{\mbox{\tiny SR}}_{\mathcal{\tilde O}}(t) \equiv -\sum\limits_{m=1}^{\tilde D} |\tilde{c}_m(t)|^2 \log |\tilde{c}_m(t)|^2,
\ee
with $\sum\limits_m |\tilde{c}_m(t)|^2 = 1$. Its upper and lower bounds are
\be
0 \le S^{\mbox{\tiny SR}}_{\mathcal{\tilde O}}(t) \le \log \tilde D.
\ee
The conjecture is that if the dynamics is chaotic within the symmetry subspace, $S^{\mbox{\tiny SR}}_{\mathcal{\tilde O}}(t)$ under most choices of operator basis should reach its maximal value in the long time limit.

To implement this concept in realistic systems, one has to overcome the hurdle of finding symmetry subspaces of the dynamics without explicit knowledge of the symmetry.
We now lay out a procedure that numerically block-diagonalizes a unitary time-evolution operator possessing robust hidden symmetries, adapted from~\cite{Chertkov2020appendixH}. 

Consider $U(\varepsilon)$, a family of unitary time-evolution operators that continuously depend on parameter $\varepsilon$. Suppose $W$ is a robust symmetry of the dynamics, namely
\be
[U(\varepsilon), W] = 0, \ \ \forall \varepsilon.
\ee
We also assume that the explicit form of $W$ is unknown. The following steps are taken in order to block-diagonalize $U(\varepsilon)$ into $W$'s symmetry subspaces:
\begin{enumerate}
    \item Choose an $\varepsilon_1$ such that the eigenstates of $U(\varepsilon_1)$, $\{ |\psi_i \rangle \}$, are non-degenerate. Under these eigenstates the alleged symmetry operator $W$ will be fully diagonalized due to $[U(\varepsilon_1), W]=0$, though the degenerate eigenvalues of $W$ (if any) will not in general be together in the diagonal form.
    
    \item Choose an $\varepsilon_2$ and evaluate all the matrix elements of 
    $ U(\varepsilon_2)$ under $\{ |\psi_i \rangle \}$. The choice of $\varepsilon_2$ can be generic as long as $U(\varepsilon_1)$ and $U(\varepsilon_2)$ do not accidentally commute. Now, as a result of $[U(\varepsilon_2),W] = 0$,
    $\langle \psi_i | U(\varepsilon_2) | \psi_j \rangle$ will be non-zero if $|\psi_i \rangle$ and $|\psi_j \rangle$ belong to the same symmetry subspace of $W$, and zero if they don't. 
    The matrix of $U(\varepsilon_2)$ under $\{ |\psi_i \rangle \}$, 
    $U(\varepsilon_2)\bigr\rvert_{\{ |\psi_i \rangle \} }$, can be viewed as an adjacency matrix describing the connectivity of an undirected graph~\cite{Chertkov2020appendixH}.

    \item One then reshuffles $\{ |\psi_i \rangle \}$ into $\{ |\tilde{\psi_i} \rangle \}$ such that $U(\varepsilon_2)\bigr\rvert_{\{ |\psi_i \rangle \} }$ becomes block-diagonal (i.e. rearranging its rows and columns). This procedure is equivalent to searching for the connected components of the undirected graph, after which each cluster of connected components forms a symmetry sector of $W$. Correspondingly, under $\{ |\tilde{\psi_i} \rangle \}$ the degenerate eigenvalues of $W$ will be grouped together, 
    \be
W\bigr\rvert_{\{ \tilde{\psi_i} \rangle  \} } =  \begin{pmatrix}
w_1 \mathbb{I}\\
&w_2 \mathbb{I}\\  
&&\ddots
\end{pmatrix}
    \ee
where $\mathbb{I}$'s are identity matrices with proper dimensions. $U(\varepsilon)$ will subsequently be block-diagonalized by $\{ |\tilde{\psi_i} \rangle \}$ for any $\varepsilon$.
\end{enumerate}
The evaluation of the symmetry-resolved entropy then follows naturally once we obtain the block structure of $U(\varepsilon)$. In general one typically performs the procedure above multiple times, zooming in on one block each time, to exhaust all the hidden commuting symmetries.

Next, we introduce the model that serves as the testing ground for our scheme, the quantum cat maps with additional symmetries.


\section{The quantum cat maps with symmetries}
\label{sec: QuantumCatMaps}
We will start by introducing the classical cat maps on a torus phase space and briefly discuss their selective key features relating to classical chaos. We then proceed to the quantum case and present the necessary ingredients required for the discussion of quantum chaos and operator growth, including the operator Hilbert space and the quantum evolution operator. We conclude the Section by presenting a class of quantum cat maps with certain symmetries that correspond to shift of momentum $p$ in the classical limit. For more thorough reviews on classical and quantum cat maps, see~\cite{Hannay-Berry, Keating1991, Chen2018}. For an in-depth discussion of quantum cat map with symmetries, see~\cite{Esposti-Winn}.

\subsection{The Classical Cat Maps}
The original cat map, first used by Vladimir Arnold as an example of a classically chaotic system, is a linear map of the unit torus $\mathbb{T}^2$ onto itself, governed by an SL$(2, \mathbb{Z})$ matrix $A = \begin{pmatrix}
A_{11} & A_{12} \\
A_{21} & A_{22} 
\end{pmatrix}$:    
\be
\begin{pmatrix}
q' \\
p' 
\end{pmatrix} = \begin{pmatrix}
A_{11} & A_{12} \\
A_{21} & A_{22} 
\end{pmatrix} 
\begin{pmatrix}
q \\
p 
\end{pmatrix}  \ \ \mbox{mod } 1
\label{eq: originalCat}
\ee
where $\begin{pmatrix}
q \\
p 
\end{pmatrix} \in \mathbb{T}^2 = \mathbb{R}^2 / \mathbb{Z}^2$.  Since $\mathrm{det} A = 1$, the map preserves area (Lebesgue measure) on the torus. Furthermore, it is required that $\mbox{Tr}A = A_{11} + A_{22} > 2 $ so that the map is hyperbolic and chaotic, and the Lyapunov exponents $\lambda_{\pm}$ are given by the logarithm of the eigenvalues of $A$, with $\lambda_{+} > 0$ being in charge of the stretching and $\lambda_{-} < 0$ the compressing directions of the hyperbolicity, respectively. 

A variation of the original cat map is to add a non-linear area-preserving perturbation:
\be
\begin{pmatrix}
q' \\
p' 
\end{pmatrix} = \begin{pmatrix}
A_{11} & A_{12} \\
A_{21} & A_{22} 
\end{pmatrix} 
\begin{pmatrix}
q \\
p 
\end{pmatrix}   +  \frac{\kappa}{2\pi} \cos (2 \pi q) \begin{pmatrix}
A_{12} \\
A_{22} 
\end{pmatrix} \ \ \mbox{mod } 1.
\label{eq: perturbedCat}
\ee
There exists a $\kappa_{\mbox{\tiny max}}$ for $\kappa$ below which the perturbed cat map is fully chaotic~\cite{Backer2003book}:
\be
\kappa_{\mbox{\tiny max}} = \frac{\sqrt{(\mbox{Tr} A)^2 - 4} - \mbox{Tr} A + 2}{2 \mbox{max}_q |H'(q)| \sqrt{1+A_{22}^2}}
\label{eq: kappaMax}
\ee
where $H(q) \equiv \frac{1}{2\pi} \cos (2\pi q)$. When $\kappa > \kappa_{\mbox{\tiny max}}$ the cat map becomes a mixed system with chaotic and non-chaotic regions coexisting in the phase space. We will limit our discussion to $\kappa \le \kappa_{\mbox{\tiny max}}$. 

Both the original and the perturbed cat maps are well understood in terms of the Lyapunov exponents, mixing behaviors, Kolmogorov-Sinai and topological entropies \cite{WaltersTextbook}. We will use them as the starting point and introduce their quantized versions, which are much less understood and will be the main focus of the rest of the paper. We will return to the connection between the classical and quantum cat maps in Section~\ref{sec: conclusion}.

\subsection{The Quantum Cat Maps}
Our main objective is the quantum counterparts of ~\eqref{eq: perturbedCat} (of which~\eqref{eq: originalCat} is a special case). Below we sketch the quantization procedure. Details can be found in~\cite{Hannay-Berry, Keating1991, Shivaji2019, Backer2003book}. 

\subsubsection{State Hilbert Space and Dynamics}
\label{subsec: SecIIIA}
We first quantize the torus. Upon quantization, $q,p$ are no longer continuous and can be viewed as the position and momentum of a single quantum particle travelling in a one-dimensional periodic lattice with $N$ sites:
\be
p = k/N,\ \  q = j/N, \ \ k,j = 1,..., N.
\ee
Furthermore, the periodicity of the wavefunction along both $p$ and $q$ directions requires that the Planck's constant $\hbar$ can only take values $\hbar = 1/(2 \pi N)$, with $N \to \infty$ corresponding to the classical limit. For the $N$ dimensional Hilbert space one can choose a set of position basis $\{ | q_j  \rangle  \}_{j = 1,..., N}$ where $q_j = j/N$ labels the discrete position.

We then quantize the classical dynamics. The quantum propogator for the classical map~\eqref{eq: perturbedCat} can be presented by an $N \times N$ unitary matrix $U$. Under the basis $\{ | q_j  \rangle  \}$, the matrix element reads
\begin{widetext}
\be
U_{kj} = \Big(\frac{A_{12}}{N} \Big)^{1/2} \exp\Bigg(\frac{i \pi}{A_{12} N}  (A_{11}j^2 - 2jk + A_{22} k^2 ) + \frac{i \kappa N }{2 \pi} \sin \Big(\frac{2 \pi j}{N}\Big)\Bigg)  G\Bigg(N' A_{11}, A'_{12}, \frac{2(A_{11}j - k)}{\mbox{gcd}(N,A_{12})}\Bigg)
\label{eq: perturbedCatU}
\ee
\end{widetext}
where $\mbox{gcd}(N,A_{12})$ is the greatest common divisor of $N$ and $A_{12}$, 
$N' \equiv N/\mbox{gcd}(N,A_{12}), \ \ A'_{12} \equiv A_{12}/\mbox{gcd}(N,A_{12})$, 
and $G$ is a number-theoretical function related to Gauss averages, defined as
\be
G(a,b,c) \equiv \lim_{M \to \infty} \frac{1}{2M} \sum\limits_{m = -M}^M \exp{\Big(\frac{\pi i}{b}(am^2+cm) \Big)}
\label{eq: GaussSum}
\ee
where $a,b,c$ are integers and $a$ and $b$ are required to be coprime. Detailed properties of $G(a,b,c)$ can be found in Appendix~\ref{app: GaussSum}.

The original and perturbed quantum cat maps studied in~\cite{Shivaji2019, Chen2018} all have $A_{11}=2, \ \ A_{12}=1, \ \ A_{21}=3, \ \ A_{22}=2$. This yields $\mbox{gcd}(N, A_{12}) = 1$ for any $N$, greatly simplifying the $G$ function. In contrast, we will study a more generic scenario where $A_{12} \ne 1$. As we will see in Section~\ref{subsec: SecIIIC} this leads to additional quantum symmetries of~\eqref{eq: perturbedCatU} under specific choices of $N$.

\subsubsection{Operator Hilbert Space and Dynamics}

To explore the quantum dynamics via the Heisenberg time evolution of operators, we introduce $N^2$-dimensional operator Hilbert space on the torus.
It is spanned by a set of orthonormal basis $\{ X^m Z^n   \}_{m,n = 0,...,(N-1)}$, with $X$ and $Z$ the position and momentum translation operators
\be
X| q_{j+1}  \rangle = |q_j \rangle, \ \ Z |p_{j+1} \rangle = |p_j \rangle
\ee
where $\{ |p_j \rangle \}_{j = 1,..., N}$ are momentum eigenstates. $X$ and $Z$ can be further written as $N\times N$ shift and clock matrices under the basis $\{ |q_j \rangle  \}_{j = 1,..., N}$:
\be
X =  \begin{pmatrix}
0 & 1  & \cdots & 0\\
\vdots & \vdots & \ddots & \vdots \\
0 & 0 & \cdots & 1 \\
1 & 0 & \cdots & 0  
\end{pmatrix},
\ \ Z =  \begin{pmatrix}
1 & 0  & \cdots & 0\\
0 & \omega & \cdots & 0 \\
\vdots & \vdots & \ddots & \vdots \\
0 & 0 & \cdots & \omega^{N-1}  
\end{pmatrix}
\label{eq: XandZ}
\ee
where $\omega \equiv e^{2 \pi i /N}$. Furthermore,  $X$ and $Z$ satisfy the following properties:
\be
X^N = Z^N = 1, \ \ XZ = \omega ZX.
\ee
Any operator $\mathcal{O}$ in the operator Hilbert space can be expanded in the following way, 
\be
\mathcal{O} = \sum\limits_{m,n=0}^{N-1} c_{mn} X^m Z^n,
\label{eq: Oinitial}
\ee
where the coefficients $\{ c_{mn}\}$ obey the normalization condition
\be
\sum\limits_{m,n} |c_{mn}|^2 = 1.
\ee
We now examine the Heisenberg time evolution of $\mathcal{O}$ under $U$. After $t$ discrete time steps $\mathcal{O}$ becomes
\be
\mathcal{O}(t) = (U^\dagger)^t \mathcal{O} U^t =  \sum\limits_{m,n} c_{mn}(t) X^m Z^n
\label{eq: Ot}
\ee
where $\{ c_{mn}(t) \}$ also satisfy the normalization condition due to the unitarity of $U$. For a given initial operator, to obtain $\{ c_{mn}(t) \}$ one simply utilizes the orthonormal condition $\mbox{Tr} (X^m Z^n X^{m'} Z^{n'}) = N^2 \delta_{mm'} \delta_{nn'}$ and the matrix representations of the operators~\eqref{eq: perturbedCatU} and~\eqref{eq: XandZ}.

\subsection{The Quantum Cat Maps with Symmetries}
\label{subsec: SecIIIC}
As mentioned in Section~\ref{subsec: SecIIIA}, the quantum cat maps used in previous studies have $A_{12} = 1$, leading to a simplification of the time evolution operator $U$. Here instead, we consider a more generic case with $A_{12} \ne 1$. The particular quantum cat map we consider has
\be
A_{11} = 14, \ \ A_{12} = 13, \ \ A_{21} = 15,\ \  A_{22} = 14,
\label{eq: OurExample}
\ee
and the corresponding time evolution operator reads, based on~\eqref{eq: perturbedCatU},
\begin{widetext}
\be
U_{kj} = \Big(\frac{13}{N} \Big)^{1/2} \exp\Bigg(\frac{i \pi}{13 N}  (14j^2 - 2jk + 14 k^2 ) + \frac{i \kappa N }{2 \pi} \sin \Big(\frac{2 \pi j}{N}\Big)\Bigg)  G\Bigg(14N', A'_{12}, \frac{2(14j - k)}{\mbox{gcd}(N,13)}\Bigg).
\label{eq: perturbedCatU14131514}
\ee
\end{widetext}
The maximally allowed value of $\kappa$ is determined as $\kappa_{\mbox{\tiny max}} \approx 0.069$ by~\eqref{eq: kappaMax}. We will take $\kappa \le 0.06$ in the remainder of the paper. 

The key feature of this quantum cat map is that it has a family of quantum symmetries for any $\kappa < \kappa_{\mbox{\tiny max}}$ when $N$ is a multiple of $A_{12} = 13$~\cite{Esposti-Winn}. (More generally, such symmetries can exist when gcd$(A_{12}, A_{22}-1) \ne 1$. For a more detailed discussion see Theorem 2 in~\cite{Esposti-Winn}.)
These symmetries are labeled by $\{ W_{r/13} \}$, where $r$ is an integer and $1 \le r < 13$. In the classical limit, $ W_{r/13} $ corresponds to translations of $p$:
\be
W^{\mbox{\tiny classical}}_{r/13} \left( \begin{array}{cc}  
q \\
p \end{array} \right)    =  \left( \begin{array}{cc}  
q \\
p +r/13\end{array} \right).
\ee
$W^{\mbox{\tiny classical}}_{r/13}$ can be quantized only if $N$ is a multiple of 13 (otherwise $p$ will not remain on the discrete lattice sites during the translation). It can be shown that when $N$ is indeed a multiple of 13, the operators $ W_{r/13}$ acts on the position basis in the following way~\cite{Esposti-Winn}:
\be
W_{r/13}  |q_j \rangle = e^{2 \pi i  \cdot j \cdot r/13}  |q_j \rangle.
\label{eq: Wsymmetry}
\ee
The matrix representation of $W_{r/13}$ under $\{ | q_j \rangle \}$ is then
\be
(W_{r/13})_{kj} = \delta_{kj} e^{2 \pi i  \cdot j \cdot r/13},
\ee
i.e. $ W_{r/13}$ is diagonal and can be rewritten in terms of $Z$ in~\eqref{eq: XandZ} as
\be
W_{r/13} = (Z)^{\frac{Nr}{13} }.
\label{eq: WusingZ}
\ee
For fixed $r$ the time evolution operator $U$ can be block-diagonalized into 13 blocks, each corresponding to a symmetry sector of $W_{r/13}$ with dimension $(N/13)^2$. Furthermore, all the values of $r$ (12 in total) yield the same block-diagonal structure of $U$.

\begin{figure}
	\begin{center}
	\includegraphics[width=2.8in]{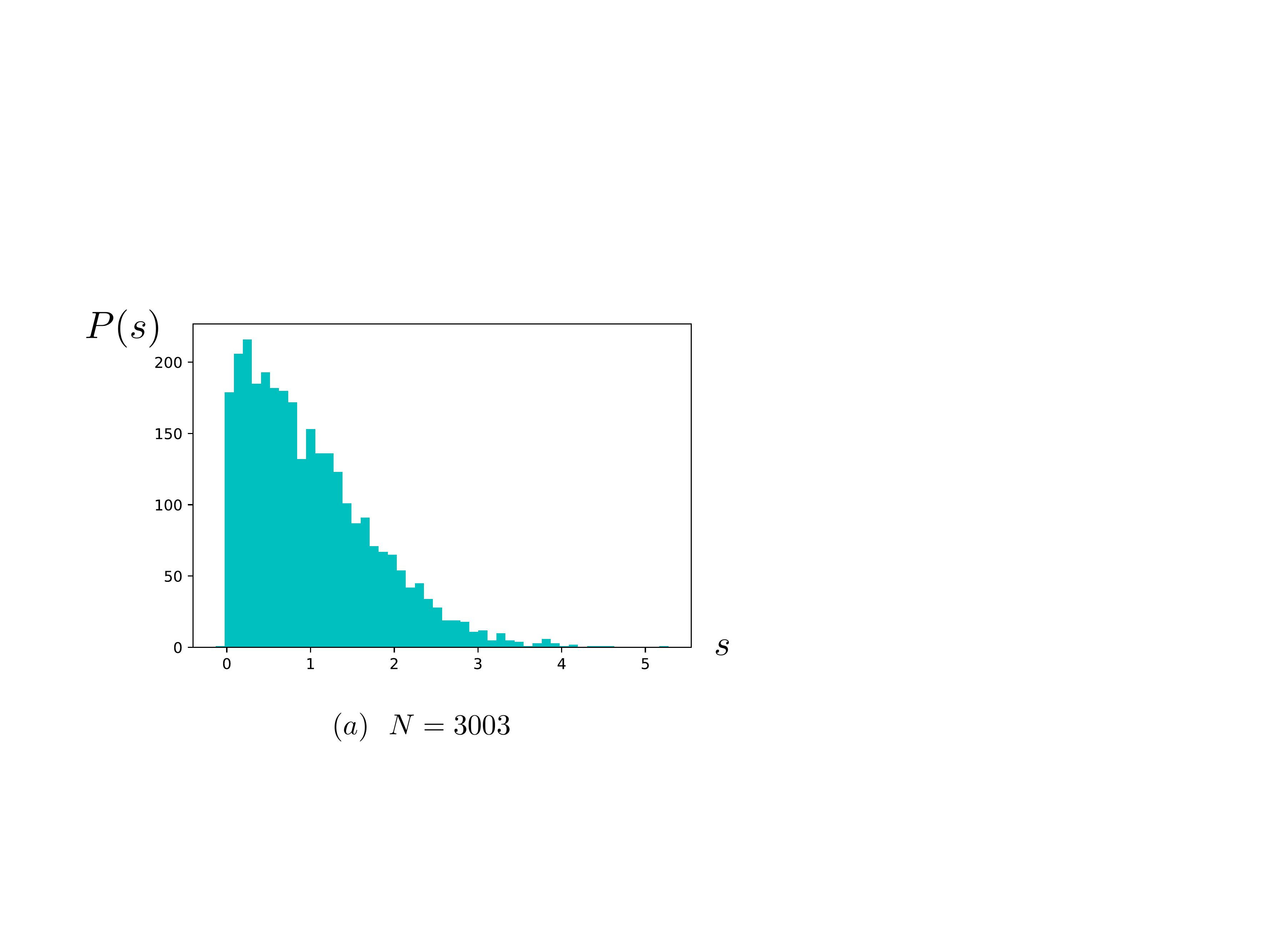}
	\includegraphics[width=2.8in]{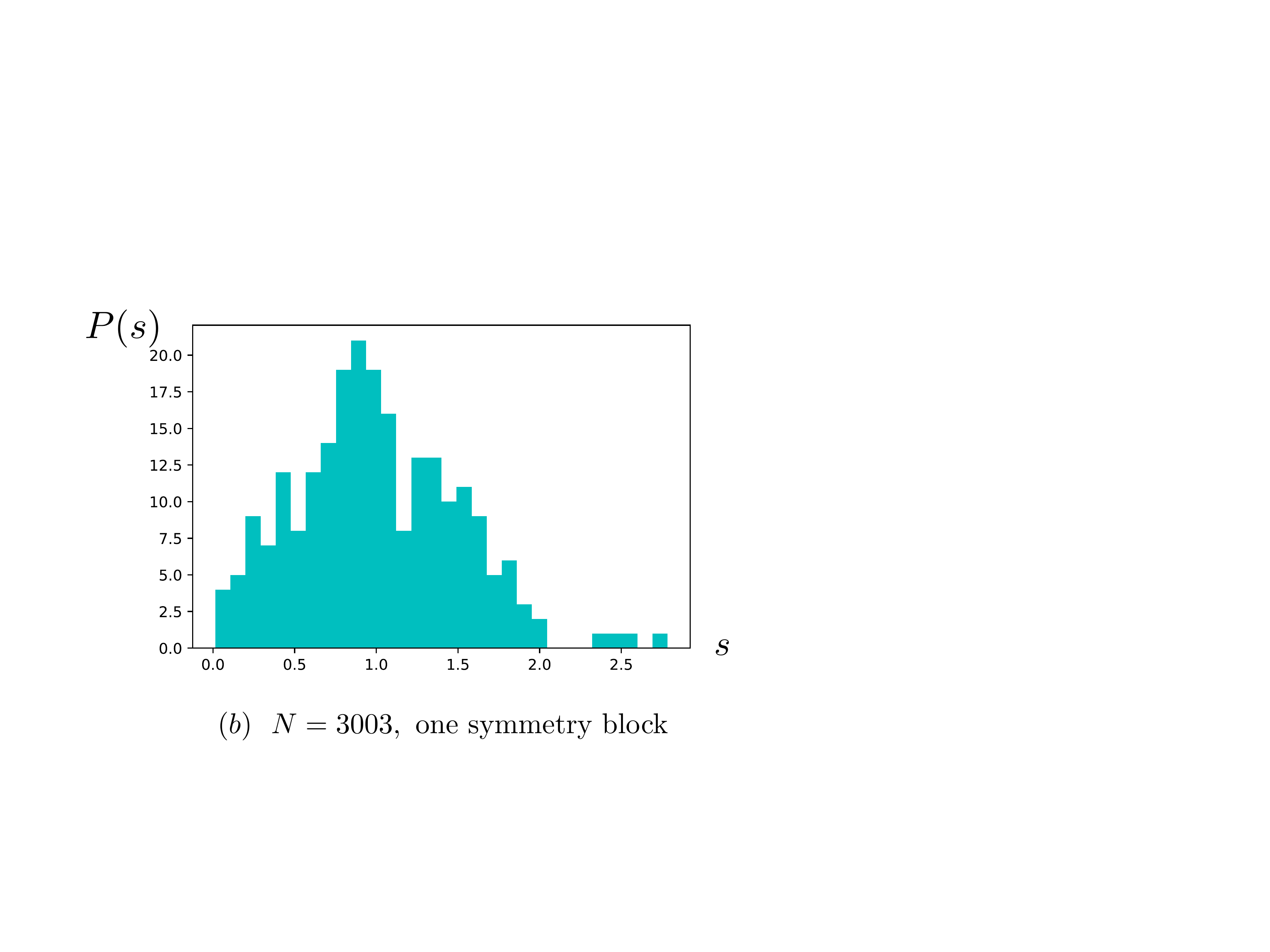}
	\includegraphics[width=2.8in]{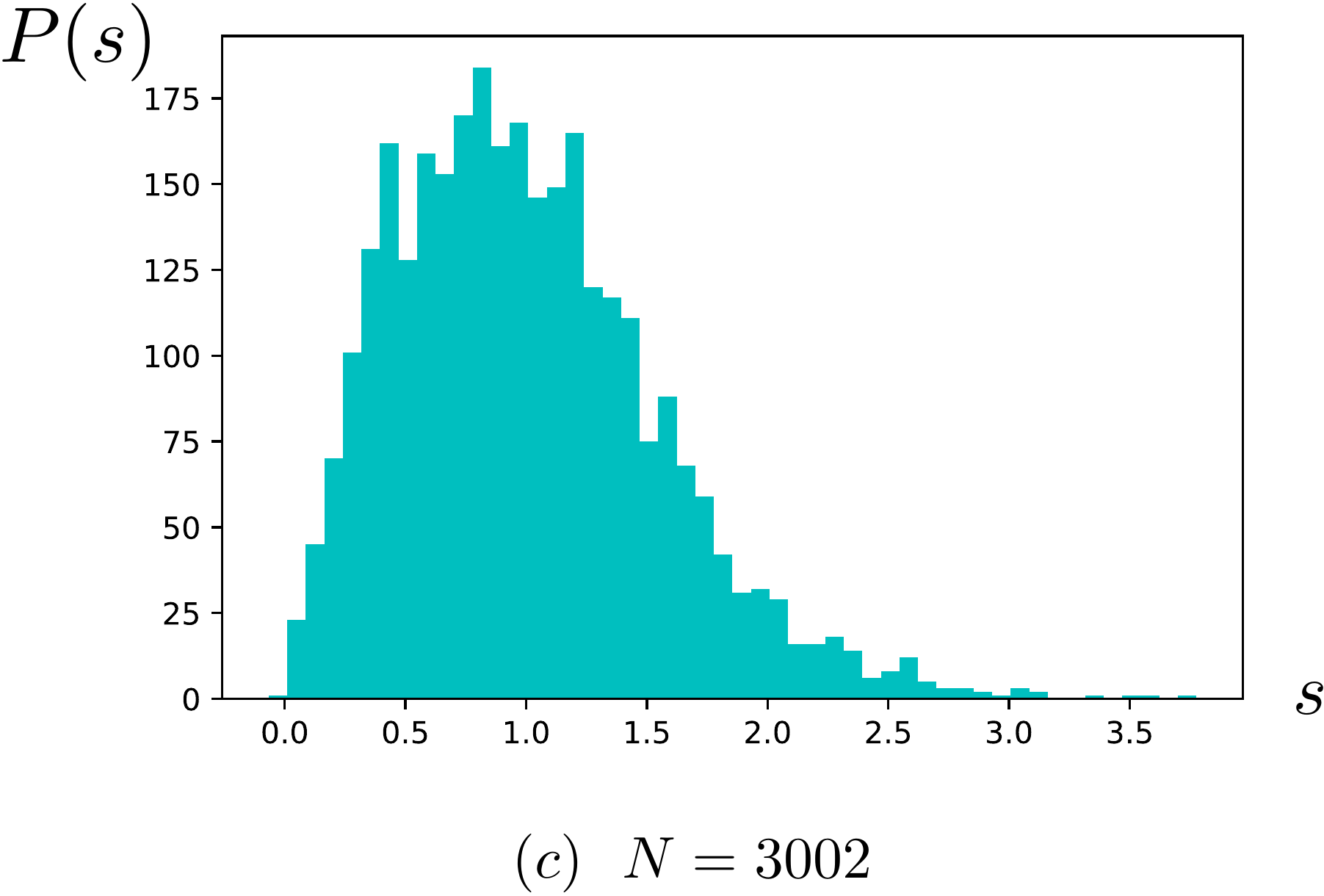}
	\end{center}
	\caption{Unfolded level statistics of quantum cat map~\eqref{eq: OurExample} with $\kappa = 0.06$. (a) $N=3003$. The Poisson-like shape is a result of sum of 13 independent Wigner-Dyson statistics. (b) $N=3003$, unfolded level statistics of one of the symmetry blocks (231 levels). The distribution shows some irregularity due to the relatively small number of levels within the block. (c) $N=3002$, where no symmetries exist.}
	\label{fig: LevelStat}
\end{figure}

The existence of $\{ W_{r/13} \}$ leads to a Poisson-like distribution of nearest-neighbor level spacings, which was demonstrated in details in~\cite{Esposti-Winn}. Here we briefly review their results.
Fig.~\ref{fig: LevelStat}(a) shows the unfolded level statistics \footnote{We perform the unfolding procedure to the energy spectrum such that the mean level spacing is one everywhere. See~\cite{Guhr1998unfolding} for details.} of the quantum cat map~\eqref{eq: OurExample} with $\kappa = 0.06$ and $N = 3003$ (which is a multiple of 13). The Poisson-like statistics is due to the superposition of levels from 13 symmetry blocks, each of which possesses Wigner-Dyson statistics as shown in Fig.~\ref{fig: LevelStat}(b). In comparison, the same cat map with $N = 3002$ exhibits Wigner-Dyson statistics due to the lack of symmetries, as shown in Fig.~\ref{fig: LevelStat}(c).  

\section{Coefficient entropy in quantum cat maps with symmetries}
\label{sec: CEinQCatMap}
Having established the fact that there are additional symmetries when $N$ is a multiple of 13 in the quantum cat map with~\eqref{eq: OurExample},
in this Section we present the results of the coefficient entropy. We will see that under a generic choice of operator basis, the coefficient entropy defined in~\eqref{eq: CoeffEntropy} saturates to the maximal value at late time, failing to capture the existence of the symmetries. Furthermore, we demonstrate both numerically and analytically that $\{ X^mZ^n \}$, previously used in~\cite{Shivaji2019} is a special choice of operator basis under which the coefficient entropy saturates to a sub-maximal value.

\subsection{Numerical Results}
\begin{figure}[h]
	\begin{center}
	\includegraphics[width=2.6in]{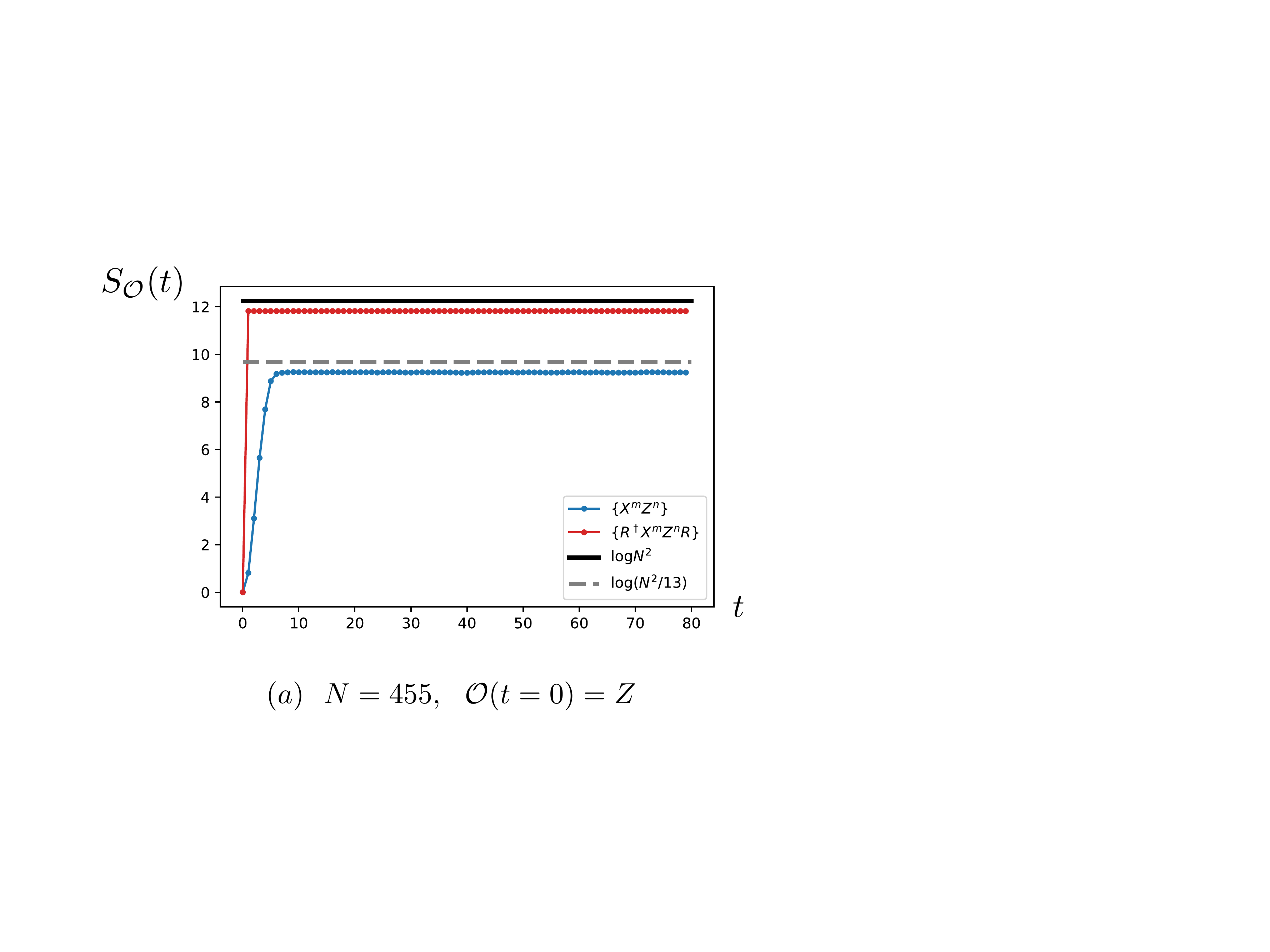}
	\includegraphics[width=2.6in]{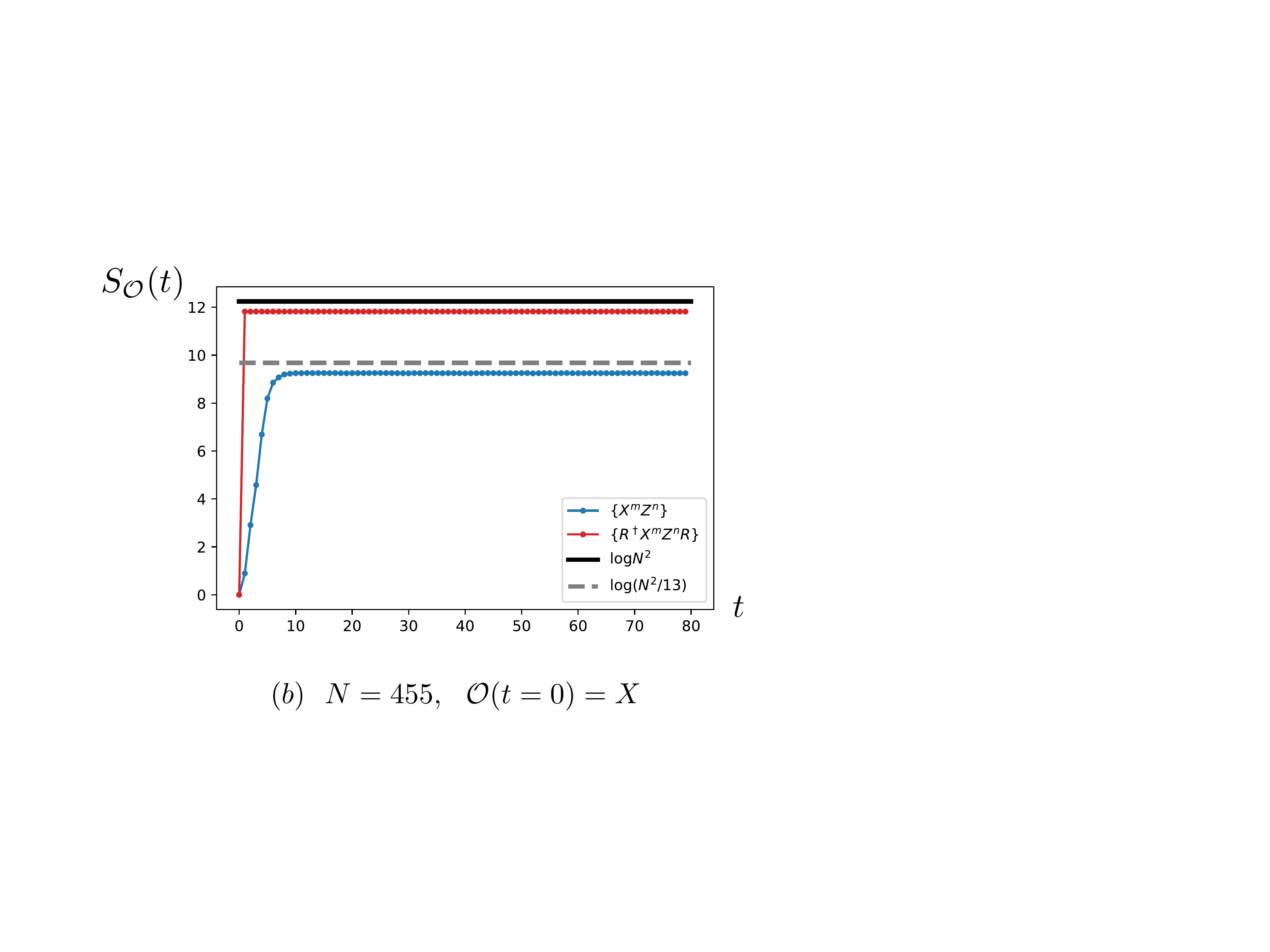}
	\includegraphics[width=2.6in]{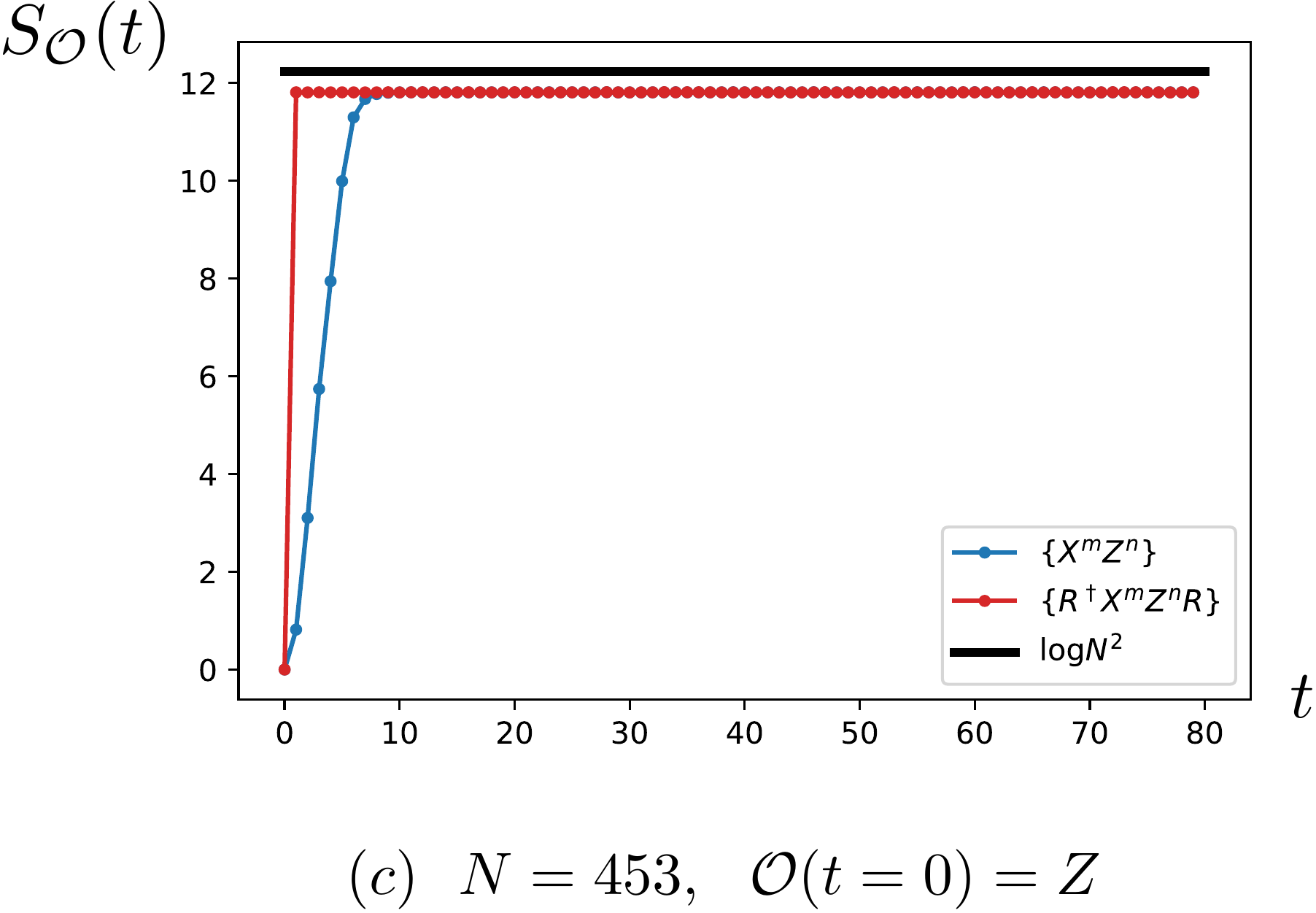}
	\end{center}
	\caption{Coefficient entropy as a function of time in quantum cat map with~\eqref{eq: OurExample} and $\kappa = 0.06$. In each plot, the blue dots are coefficient entropy data measured under the basis $\{ X^m Z^n \}$, the red dots are data measured under $\{ R^\dagger X^m Z^n  R\}$ where $R$ is a randomly chosen unitary matrix, the horizontal black line is the largest possible value $\log N^2$, and the gray dashed lines for $N=455$ mark the largest possible value $\log(N^2/13)$ in the presence of symmetries. (a) $N = 455$, and the initial operator is $Z$.
    (b) $N=455$, initial operator $X$. (c) $N=453$, which is not a multiple of 13, and initial operator $Z$. 
  }
	\label{fig: CoeffEntropy}
\end{figure}

Fig~\ref{fig: CoeffEntropy}(a) shows the coefficient entropies of quantum cat map with~\eqref{eq: OurExample} and $\kappa = 0.06$, with initial operator $Z$\footnote{Due to the complication of $A_{12} \ne 1$, we are not able to simplify the time evolution operator~\eqref{eq: perturbedCatU14131514} as was done in~\cite{Shivaji2019}, therefore are limited to relative small system size when evaluating coefficient entropy.}. Under the operator basis $\{ X^mZ^n \}$, the coefficient entropy does capture the existence of symmetries by reaching a value that is significantly smaller than $\log N^2$ at the late time. This is the coincidental circumstance we will elaborate on in the next subsection: when a ``correct" set of operator basis is selected the coefficient entropy is able to detect the hidden symmetries, and in this case $\{ X^m Z^n \}$ is such a set of basis under which the coefficient entropy saturates to $\log (N^2/13)$.
However, upon a random basis transformation corresponding to a generic choice of operator basis, the coefficient entropy becomes oblivious to the symmetries and reaches $\log N^2$. The results remain the same for a different choice of initial operator, shown in Fig~\ref{fig: CoeffEntropy}(b).
This suggests that, without the knowledge of the hidden symmetries, one generally should take great caution in using coefficient entropy to characterize operator growth. 
In comparison, if we choose $N = 453$, which is not a multiple of 13, we always obtain the same late-time saturation value (i.e. close to $\log N^2$) of the coefficient entropy regardless the choice of basis, as shown in Fig~\ref{fig: CoeffEntropy}(c).

\subsection{Analysis on the Case under $\{ X^m Z^n\}$}
We now provide analytical explanation to the sub-maximal saturation value, $\log (N^2/13)$,  of the coefficient entropy under the basis $\{ X^m Z^n \}$ when $N$ is a multiple of 13. We will show that the existence of the symmetries $\{ W_{r/13} \}_{r = 1,2,...,12}$ forces the majority of the coefficients $\{ c_{mn}(t) \}$ to be zero, suppressing the late-time coefficient entropy. For simplicity, we choose the initial operator to be $\mathcal{O}(t = 0) = Z$. The analysis for other choices of initial operators, for instance $\mathcal{O}(t=0) = X$, follows similarly. Our starting point is Eq.~\eqref{eq: Ot} which we reiterate here:
\be
(U^{\dagger})^t  Z   U^t = \sum\limits_{m,n = 0}^{N-1} c_{mn}(t) X^m Z^n.
\label{eq: UonO}
\ee
Acting $W_{r/13}^{\dagger}$ on the left and $W_{r/13}$ on the right on both sides of the above equation gives
\bea
l.h.s. &=& W^{\dagger}_{r/13} (U^{\dagger})^t Z U^t  W_{r/13} \nonumber\\
&=&   \sum\limits_{m,n} c_{mn}(t) X^m Z^n,
\label{eq: lhs}
\eea
\bea
r.h.s. &=&  \sum\limits_{m,n} c_{mn}(t)  \big[ W^{\dagger}_{r/13} X^m Z^n  W_{r/13} \big] \nonumber\\
&=& \sum\limits_{m,n} c_{mn}(t) \ \  \omega^{\frac{Nmr}{13}} X^m Z^n
\label{eq: rhs}
\eea
where we have used the following facts
\bea
&& [W_{r/13}, U] = 0, \ \ [W_{r/13}, Z] = 0, \nonumber\\
&& Z^{-\frac{Nr}{13}} X^m  = \omega^{\frac{Nmr}{13}} X^m Z^{-\frac{Nr}{13}}.
\eea
Eqs.~\eqref{eq: lhs} and~\eqref{eq: rhs} combined with the orthogonality of $\{ X^m Z^n \}$ yield
\be
c_{mn}(t) = c_{mn}(t) \omega^{\frac{Nmr}{13}} = c_{mn}(t) e^{\frac{2\pi i mr}{13}}, \ \ \forall m,n.
\ee
This means $c_{mn}(t)$ is nonzero only when $e^{2\pi i mr/13}  = 1$. Furthermore, $1 \le r < 13$ and 13 is a prime number, so $m$ needs to be a multiple of 13. Therefore, in order to have $c_{mn}(t) \ne 0$, $(m,n)$ has to satisfy:
\be
m =0,13,26, ..., (N-13), \ \ n \mbox{ arbitrary}.
\label{eq: mnconstraint}
\ee
This leads to a total $N/13 \times N = N^2 /13$ sets of nonzero $c_{mn}(t)$. The largest possible value of the coefficient entropy is then 
\be
S_{\mathcal{O}}(t \to \infty) = \log (N^2/13).
\ee
We make a few remarks here. Firstly, the analysis above is applicable to any time $t$, meaning that some $c_{mn}$'s are suppressed from the very beginning. Secondly, it is generally true that many of the $c_{mn}$'s will be constrained to be zero in the presence of the symmetries regardless of the choice of the initial operators. Our choice $\mathcal{O}(t=0) = Z$ is the simplest since $Z$ commutes with $W_{r/13}$, but one can easily show that constraint similar to~\eqref{eq: mnconstraint} emerges as long as one starts from an operator with small coefficient entropy. For example, for $\mathcal{O}(t=0) = X$, the contraint becomes
\be
m = 1,14,27, ..., (N-12), \ \ n \mbox{ arbitrary}.
\ee
Finally, one can still carry out similar analysis when $A_{12}$ is not a prime number, in which case the constraint on $(m,n)$ becomes less stringent than~\eqref{eq: mnconstraint}.



\section{Symmetry-Resolved Coefficient Entropy in Quantum Cat Maps with Symmetries}
\label{sec: SRCEinQCatMap}
\begin{figure}
	\begin{center}
	\includegraphics[width=2.6in]{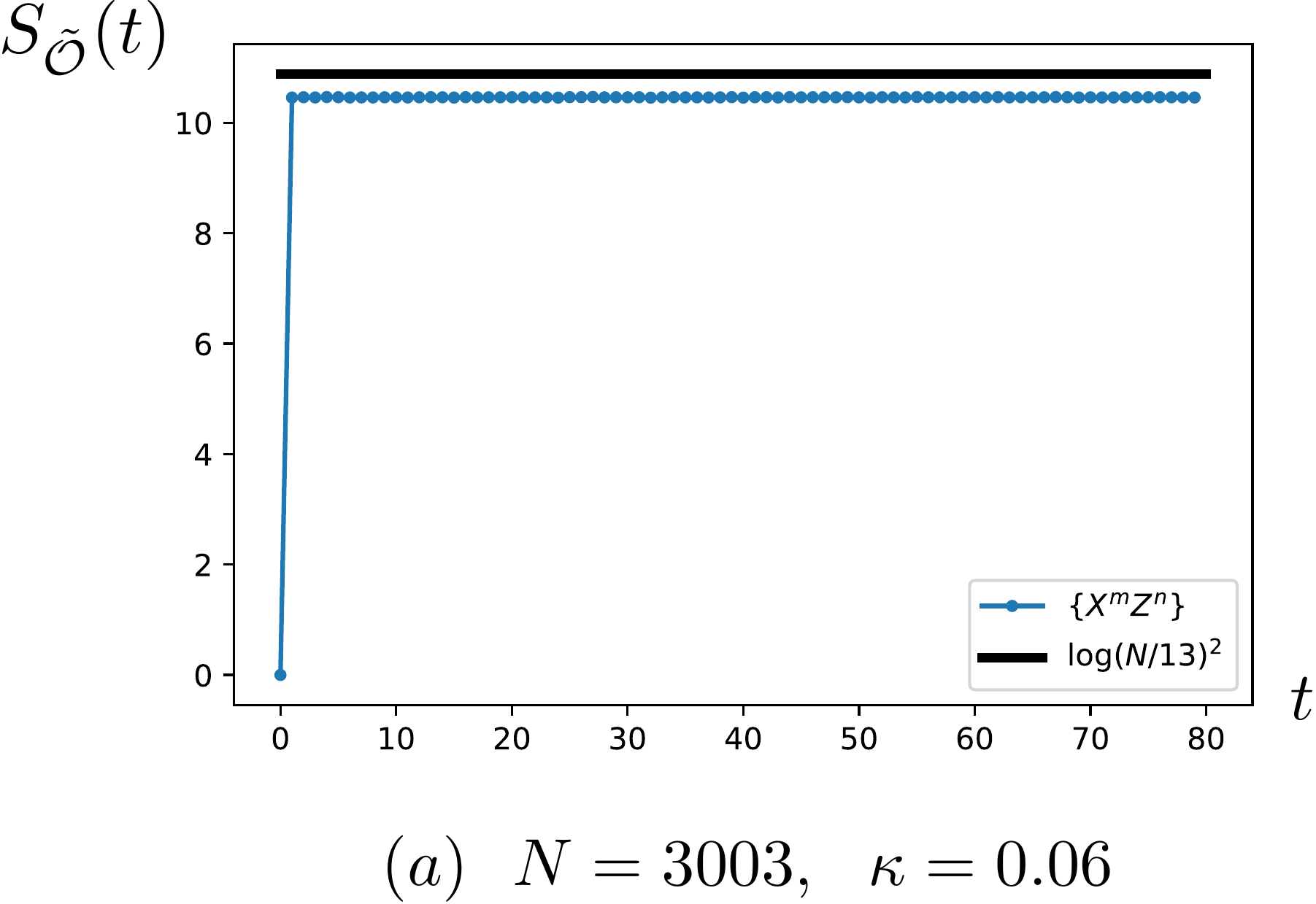}
	\includegraphics[width=2.6in]{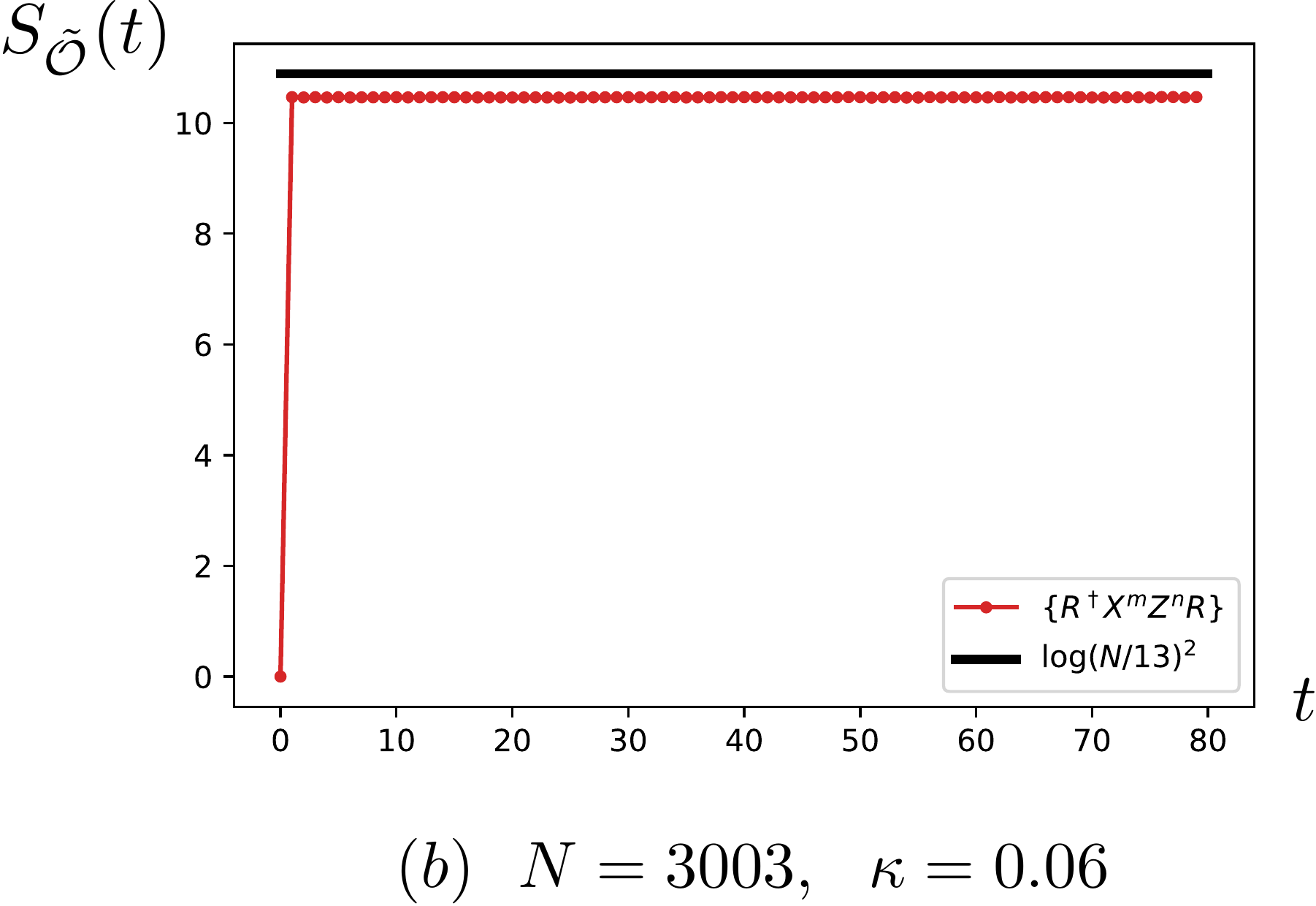}
	\includegraphics[width=2.6in]{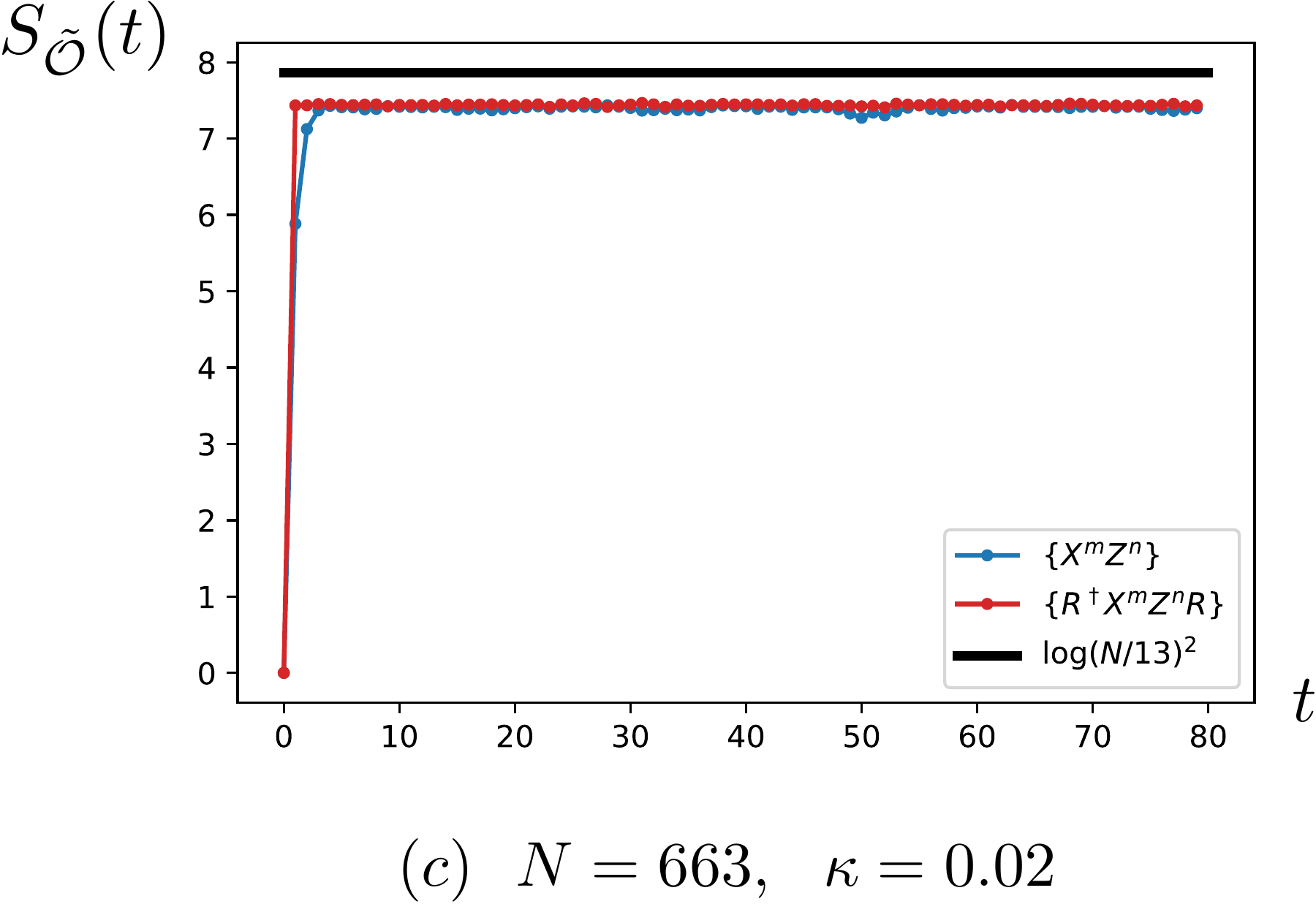}
	\includegraphics[width=2.6in]{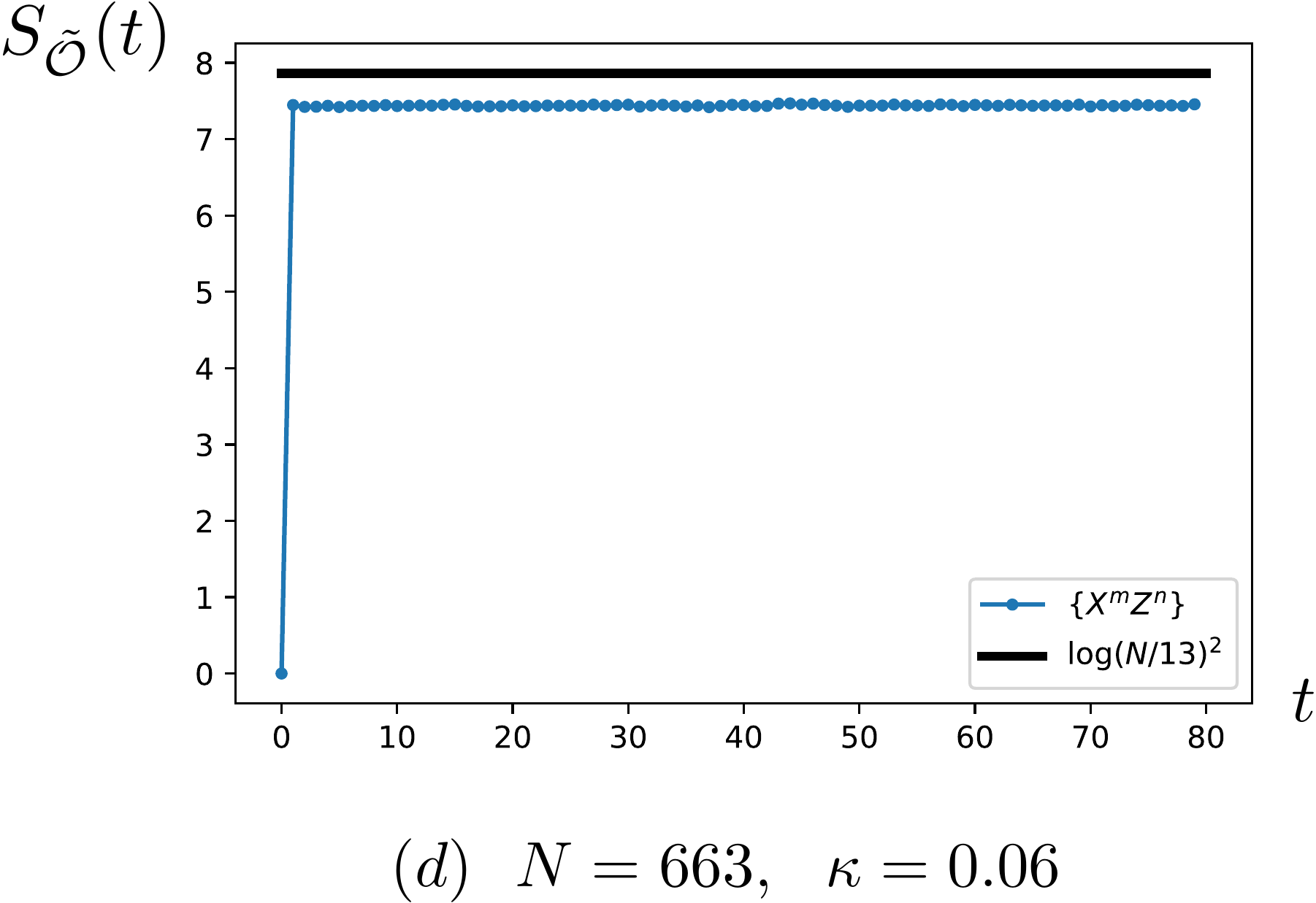}
	\end{center}
	\caption{Symmetry-resolved coefficient entropy as a function of time in quantum cat map~\eqref{eq: OurExample}, at various $N$ and $\kappa$. The size of the symmetry block is $(N/13) \times (N/13)$. The horizontal black lines in the plots represent $\log (N/13)^2$, the maximal value allowed in the symmetry subspace. (a) $N = 3003$, $\kappa = 0.06$, measured under
the operator basis $\{ X^mZ^n \}$. (b) $N = 3003$, $\kappa = 0.06$, measured under a random set of operator basis $\{ R^\dagger X^mZ^n  R \}$. (c) $N = 663$, $\kappa = 0.02$, measured under $\{ X^mZ^n \}$ and $\{ R^\dagger X^mZ^n  R \}$ where $R$ is a random unitary matrix. (d) $N = 663$, $\kappa = 0.06$, measured under $\{ X^mZ^n \}$. }
	\label{fig: SRCoeffEntropy}
\end{figure}

We now present the symmetry-resolved coefficient entropy $S_{\tilde{\mathcal{O}}}(t)$ in the cat map~\eqref{eq: OurExample} with $\kappa=0.06$. 
We pretend not knowing the explicit form of the symmetries and follow the procedure introduced in Section~\ref{subsec: SRCE}, where $\kappa$ plays the role of $\varepsilon$. 
The numerical procedure generates 13 symmetry blocks with equal sizes, consistent with the symmetries $\{ W_{r/13} \}$. 
Fig~\ref{fig: SRCoeffEntropy}(a) shows $S_{\tilde{\mathcal{O}}}(t)$ in one symmetry block of the time evolution operator with $N = 3003$, measured under the basis $\{ X^m Z^n \}$, where $X$ and $Z$ are shift and clock matrices of size $231 \times 231$. It saturates to a value that is very close to the maximal value $\log (N/13)^2$, where $(N/13)^2$ is the dimension of the symmetry block. Upon a random basis transformation, the late-time saturation value of $S_{\tilde{\mathcal{O}}}(t)$ remains unchanged, as seen in Fig~\ref{fig: SRCoeffEntropy}(b). These results do not depend on which symmetry block is chosen. Furthermore, changing the value of $\kappa$ or the initial operator does not alter the late-time saturation value of $S_{\tilde{\mathcal{O}}}(t)$. These results indicate that each symmetry block hosts maximal operator growth on its own, consistent with the observation from level statistics and from the explicit form of the symmetries.

While our main focus is the late-time saturation value of $S_{\tilde{\mathcal{O}}}(t)$, we notice
an interesting feature of the rapid growth of $S_{\tilde{\mathcal{O}}}(t)$ at early time in both Fig~\ref{fig: SRCoeffEntropy}(a) and (b): this is presumably due to the combination of a large $N$ and $\kappa$ being close to $\kappa_{\mbox{\tiny max}}$. Indeed, Fig~\ref{fig: SRCoeffEntropy}(c) shows the case with a smaller $\kappa N$ where $S_{\tilde{\mathcal{O}}}(t)$ exhibits a gradual increase when measured under $\{ X^m Z^n \}$. On the other hand, reducing $N$ alone does not seem to have an effect on the early time behavior, as shown in Fig~\ref{fig: SRCoeffEntropy}(d). It is possible that further reducing $N$ may tame the sudden jump at early time, but smaller $N$ suffers from severe finite size effect where the size of each symmetry block is too small for $S_{\tilde{\mathcal{O}}}(t)$ to exhibit a steady saturation. Another aspect of the finite-size effect is the growing discrepancy between the saturation value of $S_{\tilde{\mathcal{O}}}(t)$ and $\log (N/13)^2$ as $N$ decreases, which can be seen clearly by comparing Fig~\ref{fig: SRCoeffEntropy}(a) and (d).

\section{Conclusion and Discussion}
\label{sec: conclusion}
We have proposed the symmetry-resolved coefficient entropy as a measure of operator complexity and demonstrated its effectiveness in perturbed quantum cat maps with symmetries. In particular, we showed that even without any knowledge of the symmetries (aside from assuming they are robust and independent of the parameters of the unitary evolution) one is able to block-diagonalize the time evolution operator and focus on the operator growth within each symmetry subspace. In comparison, the original coefficient entropy is subject to the complication of basis-dependency and is not able to faithfully characterize the operator growth in the presence of hidden symmetries. Our method is readily generalizable to many-body system with robust symmetries, as well as to other measures of operator complexity that may also suffer from the problem of basis-dependency, for example the entanglement of operators, the spectrum of the operator reduced density matrix~\cite{Chen2018}, and the probability distribution of operator size~\cite{Roberts2018growth, Qi2019Measurement}. 

We would also like to make a few remarks on the cat map itself. Despite being a single-particle system, the quantum cat map has been a fruitful testing ground for several of the key concepts in (many-body) quantum chaos, such as operator complexity, level statistics, out-of-time-ordered correlator and Lyapunov exponents~\cite{Chen2018}, partially thanks to its well-understood classical limit and a well-defined quantization procedure. There are, however, open questions, in particular regarding the relation between the chaos seen in classical and quantum cat maps. For instance, on the classical side the original cat map and its perturbed version are topologically conjugate to each other \cite{WaltersTextbook}, whereas after quantization the operator growth in unperturbed quantum cat map becomes fundamentally different from the perturbed version\cite{Chen2018, Shivaji2019}. 
Though it is true that classical and (many-body) quantum chaos are likely of different nature, addressing these issues may equip us with more tools to characterize complex dynamical systems, and shed light on a more complete description of many-body quantum chaos.



\textbf{Acknowledgement} The author specially thanks Amie Wilkinson without whom this work would have never started. The author is also grateful for the discussions with Shinsei Ryu, Sanjay Moudgalya, Bryan Clark, and Eli Chertkov, and helpful comments from Fr\'ed\'eric Faure, Erez Berg, Shivaji Sondhi, Xiao Chen, and Yuhan Liu. 
In constructing the symmetry-resolved coefficient entropy we use~\cite{qosy} for finding the connected components of an adjacency matrix.
The work was partially supported by the Kadanoff Fellowship at the University of Chicago, the US National Science Foundation under Grant No. DMR-1725401 at the University of Illinois, and a fellowship at the Institute for Condensed Matter Theory. This work made use of the Illinois Campus Cluster, a computing resource operated by the Illinois Campus Cluster Program in conjunction with the National Center for Supercomputing Applications and supported by funds from the University of Illinois.

\begin{appendix}

\section{Gauss Sum}
\label{app: GaussSum}

\renewcommand{\theequation}{A.\arabic{equation}}
The function $G(a,b,c)$ that appears in the time evolution operator~\eqref{eq: perturbedCatU} is defined as (same as~\eqref{eq: GaussSum}):
\be
G(a,b,c) \equiv \lim_{M \to \infty}\frac{1}{2M} \sum\limits_{m = -M}^M \exp\Big(\frac{i \pi}{b} (am^2 + cm)  \Big)
\ee
where $a$ and $b$ are coprime integers. If $(ab+c)$ is not an even integer, then the average is equal to zero. Otherwise\cite{Hannay-Berry, Esposti-Winn}, 
\be
G(a,b,c) = P(a,b)T(a,b,c),
\label{eq: Gfunction}
\ee
where
\be
P(a,b) \equiv  \left\{ \begin{array}{cc}
\frac{1}{\sqrt{b}}   \left(\dfrac{a}{b}\right)^{\mbox{\tiny J}}
\exp \Big( \frac{-i \pi(b-1) }{4} \Big), \ \ b \mbox{ odd} \\
\frac{1}{\sqrt{b}}   \left(\dfrac{a}{b}\right)^{\mbox{\tiny J}}
\exp \Big( \frac{i \pi a}{4}\Big), \ \ b \mbox{ even} 
\end{array} \right.
\ee
\be
T(a,b,c) \equiv \left\{ \begin{array}{cc}
\exp\Big( \frac{-i \pi a}{b}(a  \mbox{\textbackslash} b)^2 (c/2)^2 \Big), \ \ ab \mbox{ even} \\
\exp\Big( \frac{-4i \pi a}{b}(4a  \mbox{\textbackslash} b)^2 c^2 \Big), \ \ ab \mbox{ odd}
\end{array} \right.
\ee
where $\left(\dfrac{a}{b}\right)^{\mbox{\tiny J}}$ is the Jacobi symbol of number theory (we have added the superscript J to distinguish it from fraction). 
It satisfies the following properties:
\be
\left(\dfrac{xy}{z}\right)^{\mbox{\tiny J}} = \left(\dfrac{x}{z}\right)^{\mbox{\tiny J}} \cdot \left(\dfrac{y}{z}\right)^{\mbox{\tiny J}}, \nonumber\\
\ee
\be
\left(\dfrac{x}{z}\right)^{\mbox{\tiny J}} = \left(\dfrac{y}{z}\right)^{\mbox{\tiny J}}, \ \ \mbox{if } x \equiv y \ \ (\mbox{mod }z).
\label{eq: JacobiProperty}
\ee
For small $x$ and $z$, the value of their Jacobi symbol $\left(\dfrac{x}{z}\right)^{\mbox{\tiny J}}$ can be looked up in \cite{IrelandBook}.
The notation $(a  \mbox{\textbackslash} b)$ represents an integer that is between $1$ and $(b-1)$ satisfying:
\be
a \cdot (a  \mbox{\textbackslash} b) \equiv 1 \ \  (\mbox{mod } b).
\ee
$ (a  \mbox{\textbackslash} b)$ is uniquely determined because $a$ and $b$ are coprime.
It is also related to the Euler totient function $\phi_E(b)$:
\be
(a  \mbox{\textbackslash} b) \equiv a^{\phi_E(b) - 1}    \ \ (\mbox{mod } b),
\ee
where $\phi_E(b)$ counts the positive integers up to $b$ that are relatively prime to $b$. In the case of $b = 1$, $\phi_E(1) = 1$ so $(a  \mbox{\textbackslash} b) = 0$.

In our case, as written in~\eqref{eq: perturbedCatU14131514}, 
\be
a = 14N' , \ \ b = A'_{12}, \ \ c = \frac{2(14j-k)}{\mbox{gcd}(N,13)}
\ee
where $N' \equiv N/\mbox{gcd}(N,13)$, $A'_{12} \equiv A_{12}/\mbox{gcd}(N,13)$. Take $N = 3003 = 231 \times 13$ as an example. We then have
\be
G(a,b,c) = G \Big(14 \times 231, 1, \frac{2(14j-k)}{13} \Big)
\ee
Therefore, $G(a,b,c) = 0$ if $(14j - k)$ is not a multiple of $13$. When $(14j - k)$ is a multiple of 13, $ab+c$ is automatically an even integer. Thus we have
\bea
&& G(a,b,c) = G \Big(14 \times 231, 1, \frac{2(14j-k)}{13} \Big) \nonumber\\
&=&  \delta(14j - k = 13\cdot \mbox{integer}) \cdot \left(\dfrac{14\times 231}{13}\right)^{\mbox{\tiny J}}  \nonumber\\
&& \cdot \exp\Big(-i \pi \times 14 \times 231 \times (14\times 231 \setminus 1)^2 (\mbox{integer})^2 \Big) \nonumber\\
&&
\eea
where the Jacobi symbol can be simplified using~\eqref{eq: JacobiProperty}, and $(14\times 231 \setminus 1) = 0$. Thus 
\be
G(a,b,c) = \delta(14j - k = 13\cdot \mbox{integer}).
\ee
For $N=3002$, there is no longer a $\delta$-function in $G$ because $\mbox{gcd}(N, 13) = 1$. The exponential term in $T(a,b,c)$ also becomes more complicated due to $(a  \mbox{\textbackslash} b) \ne 0$.

\end{appendix}

\bibliography{References}

\begin{thebibliography}{38}%
\makeatletter
\providecommand \@ifxundefined [1]{%
 \@ifx{#1\undefined}
}%
\providecommand \@ifnum [1]{%
 \ifnum #1\expandafter \@firstoftwo
 \else \expandafter \@secondoftwo
 \fi
}%
\providecommand \@ifx [1]{%
 \ifx #1\expandafter \@firstoftwo
 \else \expandafter \@secondoftwo
 \fi
}%
\providecommand \natexlab [1]{#1}%
\providecommand \enquote  [1]{``#1''}%
\providecommand \bibnamefont  [1]{#1}%
\providecommand \bibfnamefont [1]{#1}%
\providecommand \citenamefont [1]{#1}%
\providecommand \href@noop [0]{\@secondoftwo}%
\providecommand \href [0]{\begingroup \@sanitize@url \@href}%
\providecommand \@href[1]{\@@startlink{#1}\@@href}%
\providecommand \@@href[1]{\endgroup#1\@@endlink}%
\providecommand \@sanitize@url [0]{\catcode `\\12\catcode `\$12\catcode
  `\&12\catcode `\#12\catcode `\^12\catcode `\_12\catcode `\%12\relax}%
\providecommand \@@startlink[1]{}%
\providecommand \@@endlink[0]{}%
\providecommand \url  [0]{\begingroup\@sanitize@url \@url }%
\providecommand \@url [1]{\endgroup\@href {#1}{\urlprefix }}%
\providecommand \urlprefix  [0]{URL }%
\providecommand \Eprint [0]{\href }%
\providecommand \doibase [0]{http://dx.doi.org/}%
\providecommand \selectlanguage [0]{\@gobble}%
\providecommand \bibinfo  [0]{\@secondoftwo}%
\providecommand \bibfield  [0]{\@secondoftwo}%
\providecommand \translation [1]{[#1]}%
\providecommand \BibitemOpen [0]{}%
\providecommand \bibitemStop [0]{}%
\providecommand \bibitemNoStop [0]{.\EOS\space}%
\providecommand \EOS [0]{\spacefactor3000\relax}%
\providecommand \BibitemShut  [1]{\csname bibitem#1\endcsname}%
\let\auto@bib@innerbib\@empty
\bibitem [{\citenamefont {Moudgalya}\ \emph {et~al.}(2019)\citenamefont
  {Moudgalya}, \citenamefont {Devakul}, \citenamefont {von Keyserlingk},\ and\
  \citenamefont {Sondhi}}]{Shivaji2019}%
  \BibitemOpen
  \bibfield  {author} {\bibinfo {author} {\bibfnamefont {S.}~\bibnamefont
  {Moudgalya}}, \bibinfo {author} {\bibfnamefont {T.}~\bibnamefont {Devakul}},
  \bibinfo {author} {\bibfnamefont {C.~W.}\ \bibnamefont {von Keyserlingk}}, \
  and\ \bibinfo {author} {\bibfnamefont {S.~L.}\ \bibnamefont {Sondhi}},\
  }\href {\doibase 10.1103/PhysRevB.99.094312} {\bibfield  {journal} {\bibinfo
  {journal} {Physical Review B}\ }\textbf {\bibinfo {volume} {99}},\ \bibinfo
  {pages} {094312} (\bibinfo {year} {2019})}\BibitemShut {NoStop}%
\bibitem [{\citenamefont {Deutsch}(1991)}]{Deutsch1991}%
  \BibitemOpen
  \bibfield  {author} {\bibinfo {author} {\bibfnamefont {J.~M.}\ \bibnamefont
  {Deutsch}},\ }\href {\doibase 10.1103/PhysRevA.43.2046} {\bibfield  {journal}
  {\bibinfo  {journal} {Physical Review A}\ }\textbf {\bibinfo {volume} {43}},\
  \bibinfo {pages} {2046} (\bibinfo {year} {1991})}\BibitemShut {NoStop}%
\bibitem [{\citenamefont {{Srednicki}}(1994)}]{Srednicki1994}%
  \BibitemOpen
  \bibfield  {author} {\bibinfo {author} {\bibfnamefont {M.}~\bibnamefont
  {{Srednicki}}},\ }\href {\doibase 10.1103/PhysRevE.50.888} {\bibfield
  {journal} {\bibinfo  {journal} {Physical Review E}\ }\textbf {\bibinfo
  {volume} {50}},\ \bibinfo {pages} {888} (\bibinfo {year} {1994})}\BibitemShut
  {NoStop}%
\bibitem [{\citenamefont {Basko}\ \emph {et~al.}(2006)\citenamefont {Basko},
  \citenamefont {Aleiner},\ and\ \citenamefont {Altshuler}}]{Basko2006MBL}%
  \BibitemOpen
  \bibfield  {author} {\bibinfo {author} {\bibfnamefont {D.}~\bibnamefont
  {Basko}}, \bibinfo {author} {\bibfnamefont {I.}~\bibnamefont {Aleiner}}, \
  and\ \bibinfo {author} {\bibfnamefont {B.}~\bibnamefont {Altshuler}},\ }\href
  {\doibase https://doi.org/10.1016/j.aop.2005.11.014} {\bibfield  {journal}
  {\bibinfo  {journal} {Annals of Physics}\ }\textbf {\bibinfo {volume}
  {321}},\ \bibinfo {pages} {1126 } (\bibinfo {year} {2006})}\BibitemShut
  {NoStop}%
\bibitem [{\citenamefont {Pal}\ and\ \citenamefont
  {Huse}(2010)}]{PalHuse2010MBL}%
  \BibitemOpen
  \bibfield  {author} {\bibinfo {author} {\bibfnamefont {A.}~\bibnamefont
  {Pal}}\ and\ \bibinfo {author} {\bibfnamefont {D.~A.}\ \bibnamefont {Huse}},\
  }\href {\doibase 10.1103/PhysRevB.82.174411} {\bibfield  {journal} {\bibinfo
  {journal} {Physical Review B}\ }\textbf {\bibinfo {volume} {82}},\ \bibinfo
  {pages} {174411} (\bibinfo {year} {2010})}\BibitemShut {NoStop}%
\bibitem [{\citenamefont {{Roberts}}\ and\ \citenamefont
  {{Stanford}}(2015)}]{RobertsStanford2015diagnosing}%
  \BibitemOpen
  \bibfield  {author} {\bibinfo {author} {\bibfnamefont {D.~A.}\ \bibnamefont
  {{Roberts}}}\ and\ \bibinfo {author} {\bibfnamefont {D.}~\bibnamefont
  {{Stanford}}},\ }\href {\doibase 10.1103/PhysRevLett.115.131603} {\bibfield
  {journal} {\bibinfo  {journal} {Physical Review Letters}\ }\textbf {\bibinfo
  {volume} {115}},\ \bibinfo {eid} {131603} (\bibinfo {year}
  {2015})}\BibitemShut {NoStop}%
\bibitem [{\citenamefont {{Shenker}}\ and\ \citenamefont
  {{Stanford}}(2014)}]{ShenkerStanford2014butterfly}%
  \BibitemOpen
  \bibfield  {author} {\bibinfo {author} {\bibfnamefont {S.~H.}\ \bibnamefont
  {{Shenker}}}\ and\ \bibinfo {author} {\bibfnamefont {D.}~\bibnamefont
  {{Stanford}}},\ }\href {\doibase 10.1007/JHEP03(2014)067} {\bibfield
  {journal} {\bibinfo  {journal} {Journal of High Energy Physics}\ }\textbf
  {\bibinfo {volume} {3}},\ \bibinfo {eid} {67} (\bibinfo {year}
  {2014})}\BibitemShut {NoStop}%
\bibitem [{\citenamefont {{Maldacena}}\ \emph {et~al.}(2016)\citenamefont
  {{Maldacena}}, \citenamefont {{Shenker}},\ and\ \citenamefont
  {{Stanford}}}]{Maldacena2016bound}%
  \BibitemOpen
  \bibfield  {author} {\bibinfo {author} {\bibfnamefont {J.}~\bibnamefont
  {{Maldacena}}}, \bibinfo {author} {\bibfnamefont {S.~H.}\ \bibnamefont
  {{Shenker}}}, \ and\ \bibinfo {author} {\bibfnamefont {D.}~\bibnamefont
  {{Stanford}}},\ }\href {\doibase 10.1007/JHEP08(2016)106} {\bibfield
  {journal} {\bibinfo  {journal} {Journal of High Energy Physics}\ }\textbf
  {\bibinfo {volume} {8}},\ \bibinfo {eid} {106} (\bibinfo {year}
  {2016})}\BibitemShut {NoStop}%
\bibitem [{\citenamefont {{Cotler}}\ \emph {et~al.}(2017)\citenamefont
  {{Cotler}}, \citenamefont {{Gur-Ari}}, \citenamefont {{Hanada}},
  \citenamefont {{Polchinski}}, \citenamefont {{Saad}}, \citenamefont
  {{Shenker}}, \citenamefont {{Stanford}}, \citenamefont {{Streicher}},\ and\
  \citenamefont {{Tezuka}}}]{Cotler2017RMT}%
  \BibitemOpen
  \bibfield  {author} {\bibinfo {author} {\bibfnamefont {J.~S.}\ \bibnamefont
  {{Cotler}}}, \bibinfo {author} {\bibfnamefont {G.}~\bibnamefont {{Gur-Ari}}},
  \bibinfo {author} {\bibfnamefont {M.}~\bibnamefont {{Hanada}}}, \bibinfo
  {author} {\bibfnamefont {J.}~\bibnamefont {{Polchinski}}}, \bibinfo {author}
  {\bibfnamefont {P.}~\bibnamefont {{Saad}}}, \bibinfo {author} {\bibfnamefont
  {S.~H.}\ \bibnamefont {{Shenker}}}, \bibinfo {author} {\bibfnamefont
  {D.}~\bibnamefont {{Stanford}}}, \bibinfo {author} {\bibfnamefont
  {A.}~\bibnamefont {{Streicher}}}, \ and\ \bibinfo {author} {\bibfnamefont
  {M.}~\bibnamefont {{Tezuka}}},\ }\href {\doibase 10.1007/JHEP05(2017)118}
  {\bibfield  {journal} {\bibinfo  {journal} {Journal of High Energy Physics}\
  }\textbf {\bibinfo {volume} {5}},\ \bibinfo {eid} {118} (\bibinfo {year}
  {2017})}\BibitemShut {NoStop}%
\bibitem [{\citenamefont {{Larkin}}\ and\ \citenamefont
  {{Ovchinnikov}}(1969)}]{Larkin1969}%
  \BibitemOpen
  \bibfield  {author} {\bibinfo {author} {\bibfnamefont {A.~I.}\ \bibnamefont
  {{Larkin}}}\ and\ \bibinfo {author} {\bibfnamefont {Y.~N.}\ \bibnamefont
  {{Ovchinnikov}}},\ }\href@noop {} {\bibfield  {journal} {\bibinfo  {journal}
  {Soviet Journal of Experimental and Theoretical Physics}\ }\textbf {\bibinfo
  {volume} {28}},\ \bibinfo {pages} {1200} (\bibinfo {year}
  {1969})}\BibitemShut {NoStop}%
\bibitem [{\citenamefont {Dubail}(2017)}]{Dubail2017OpEE}%
  \BibitemOpen
  \bibfield  {author} {\bibinfo {author} {\bibfnamefont {J.}~\bibnamefont
  {Dubail}},\ }\href@noop {} {\bibfield  {journal} {\bibinfo  {journal}
  {Journal of Physics A: Mathematical and Theoretical}\ }\textbf {\bibinfo
  {volume} {50}},\ \bibinfo {pages} {234001} (\bibinfo {year}
  {2017})}\BibitemShut {NoStop}%
\bibitem [{\citenamefont {Chen}\ and\ \citenamefont {Zhou}(2018)}]{Chen2018}%
  \BibitemOpen
  \bibfield  {author} {\bibinfo {author} {\bibfnamefont {X.}~\bibnamefont
  {Chen}}\ and\ \bibinfo {author} {\bibfnamefont {T.}~\bibnamefont {Zhou}},\
  }\href@noop {} {\enquote {\bibinfo {title} {Operator scrambling and quantum
  chaos},}\ } (\bibinfo {year} {2018}),\ \Eprint
  {http://arxiv.org/abs/1804.08655} {arXiv:1804.08655 [cond-mat.str-el]}
  \BibitemShut {NoStop}%
\bibitem [{\citenamefont {Kudler-Flam}\ \emph {et~al.}(2021)\citenamefont
  {Kudler-Flam}, \citenamefont {Nozaki}, \citenamefont {Ryu},\ and\
  \citenamefont {Tan}}]{Kudler-Flam2021local}%
  \BibitemOpen
  \bibfield  {author} {\bibinfo {author} {\bibfnamefont {J.}~\bibnamefont
  {Kudler-Flam}}, \bibinfo {author} {\bibfnamefont {M.}~\bibnamefont {Nozaki}},
  \bibinfo {author} {\bibfnamefont {S.}~\bibnamefont {Ryu}}, \ and\ \bibinfo
  {author} {\bibfnamefont {M.~T.}\ \bibnamefont {Tan}},\ }\href {\doibase
  10.1103/PhysRevResearch.3.033182} {\bibfield  {journal} {\bibinfo  {journal}
  {Physical Review Research}\ }\textbf {\bibinfo {volume} {3}},\ \bibinfo
  {pages} {033182} (\bibinfo {year} {2021})}\BibitemShut {NoStop}%
\bibitem [{\citenamefont {Parker}\ \emph {et~al.}(2019)\citenamefont {Parker},
  \citenamefont {Cao}, \citenamefont {Avdoshkin}, \citenamefont {Scaffidi},\
  and\ \citenamefont {Altman}}]{Parker2019Krylov}%
  \BibitemOpen
  \bibfield  {author} {\bibinfo {author} {\bibfnamefont {D.~E.}\ \bibnamefont
  {Parker}}, \bibinfo {author} {\bibfnamefont {X.}~\bibnamefont {Cao}},
  \bibinfo {author} {\bibfnamefont {A.}~\bibnamefont {Avdoshkin}}, \bibinfo
  {author} {\bibfnamefont {T.}~\bibnamefont {Scaffidi}}, \ and\ \bibinfo
  {author} {\bibfnamefont {E.}~\bibnamefont {Altman}},\ }\href {\doibase
  10.1103/PhysRevX.9.041017} {\bibfield  {journal} {\bibinfo  {journal}
  {Physical Review X}\ }\textbf {\bibinfo {volume} {9}},\ \bibinfo {pages}
  {041017} (\bibinfo {year} {2019})}\BibitemShut {NoStop}%
\bibitem [{\citenamefont {Kar}\ \emph {et~al.}(2021)\citenamefont {Kar},
  \citenamefont {Lamprou}, \citenamefont {Rozali},\ and\ \citenamefont
  {Sully}}]{Kar2021KrylovGravity}%
  \BibitemOpen
  \bibfield  {author} {\bibinfo {author} {\bibfnamefont {A.}~\bibnamefont
  {Kar}}, \bibinfo {author} {\bibfnamefont {L.}~\bibnamefont {Lamprou}},
  \bibinfo {author} {\bibfnamefont {M.}~\bibnamefont {Rozali}}, \ and\ \bibinfo
  {author} {\bibfnamefont {J.}~\bibnamefont {Sully}},\ }\href@noop {} {\enquote
  {\bibinfo {title} {Random matrix theory for complexity growth and black hole
  interiors},}\ } (\bibinfo {year} {2021}),\ \Eprint
  {http://arxiv.org/abs/2106.02046} {arXiv:2106.02046 [hep-th]} \BibitemShut
  {NoStop}%
\bibitem [{\citenamefont {Caputa}\ and\ \citenamefont
  {Datta}(2021)}]{Caputa2021KrylovCFT}%
  \BibitemOpen
  \bibfield  {author} {\bibinfo {author} {\bibfnamefont {P.}~\bibnamefont
  {Caputa}}\ and\ \bibinfo {author} {\bibfnamefont {S.}~\bibnamefont {Datta}},\
  }\href@noop {} {\enquote {\bibinfo {title} {Operator growth in 2d cft},}\ }
  (\bibinfo {year} {2021}),\ \Eprint {http://arxiv.org/abs/2110.10519}
  {arXiv:2110.10519 [hep-th]} \BibitemShut {NoStop}%
\bibitem [{\citenamefont {Roberts}\ \emph {et~al.}(2018)\citenamefont
  {Roberts}, \citenamefont {Stanford},\ and\ \citenamefont
  {Streicher}}]{Roberts2018growth}%
  \BibitemOpen
  \bibfield  {author} {\bibinfo {author} {\bibfnamefont {D.~A.}\ \bibnamefont
  {Roberts}}, \bibinfo {author} {\bibfnamefont {D.}~\bibnamefont {Stanford}}, \
  and\ \bibinfo {author} {\bibfnamefont {A.}~\bibnamefont {Streicher}},\
  }\href@noop {} {\bibfield  {journal} {\bibinfo  {journal} {Journal of High
  Energy Physics}\ }\textbf {\bibinfo {volume} {2018}},\ \bibinfo {pages} {122}
  (\bibinfo {year} {2018})}\BibitemShut {NoStop}%
\bibitem [{\citenamefont {Qi}\ \emph {et~al.}(2019)\citenamefont {Qi},
  \citenamefont {Davis}, \citenamefont {Periwal},\ and\ \citenamefont
  {Schleier-Smith}}]{Qi2019Measurement}%
  \BibitemOpen
  \bibfield  {author} {\bibinfo {author} {\bibfnamefont {X.-L.}\ \bibnamefont
  {Qi}}, \bibinfo {author} {\bibfnamefont {E.~J.}\ \bibnamefont {Davis}},
  \bibinfo {author} {\bibfnamefont {A.}~\bibnamefont {Periwal}}, \ and\
  \bibinfo {author} {\bibfnamefont {M.}~\bibnamefont {Schleier-Smith}},\
  }\href@noop {} {\  (\bibinfo {year} {2019})},\ \Eprint
  {http://arxiv.org/abs/1906.00524} {arXiv:1906.00524 [quant-ph]} \BibitemShut
  {NoStop}%
\bibitem [{\citenamefont {Mondaini}\ \emph {et~al.}(2016)\citenamefont
  {Mondaini}, \citenamefont {Fratus}, \citenamefont {Srednicki},\ and\
  \citenamefont {Rigol}}]{Mondaini2016IsingRMT}%
  \BibitemOpen
  \bibfield  {author} {\bibinfo {author} {\bibfnamefont {R.}~\bibnamefont
  {Mondaini}}, \bibinfo {author} {\bibfnamefont {K.~R.}\ \bibnamefont
  {Fratus}}, \bibinfo {author} {\bibfnamefont {M.}~\bibnamefont {Srednicki}}, \
  and\ \bibinfo {author} {\bibfnamefont {M.}~\bibnamefont {Rigol}},\ }\href
  {\doibase 10.1103/PhysRevE.93.032104} {\bibfield  {journal} {\bibinfo
  {journal} {Physical Review E}\ }\textbf {\bibinfo {volume} {93}},\ \bibinfo
  {pages} {032104} (\bibinfo {year} {2016})}\BibitemShut {NoStop}%
\bibitem [{\citenamefont {{Balasubramanian}}\ \emph {et~al.}(2017)\citenamefont
  {{Balasubramanian}}, \citenamefont {{Craps}}, \citenamefont {{Czech}},\ and\
  \citenamefont {{S{\'a}rosi}}}]{Balasubramanian2017echos}%
  \BibitemOpen
  \bibfield  {author} {\bibinfo {author} {\bibfnamefont {V.}~\bibnamefont
  {{Balasubramanian}}}, \bibinfo {author} {\bibfnamefont {B.}~\bibnamefont
  {{Craps}}}, \bibinfo {author} {\bibfnamefont {B.}~\bibnamefont {{Czech}}}, \
  and\ \bibinfo {author} {\bibfnamefont {G.}~\bibnamefont {{S{\'a}rosi}}},\
  }\href {\doibase 10.1007/JHEP03(2017)154} {\bibfield  {journal} {\bibinfo
  {journal} {Journal of High Energy Physics}\ }\textbf {\bibinfo {volume}
  {2017}},\ \bibinfo {eid} {154} (\bibinfo {year} {2017})}\BibitemShut
  {NoStop}%
\bibitem [{\citenamefont {Cotler}\ \emph {et~al.}(2017)\citenamefont {Cotler},
  \citenamefont {Hunter-Jones}, \citenamefont {Liu},\ and\ \citenamefont
  {Yoshida}}]{Cotler2017complexity}%
  \BibitemOpen
  \bibfield  {author} {\bibinfo {author} {\bibfnamefont {J.}~\bibnamefont
  {Cotler}}, \bibinfo {author} {\bibfnamefont {N.}~\bibnamefont
  {Hunter-Jones}}, \bibinfo {author} {\bibfnamefont {J.}~\bibnamefont {Liu}}, \
  and\ \bibinfo {author} {\bibfnamefont {B.}~\bibnamefont {Yoshida}},\
  }\href@noop {} {\bibfield  {journal} {\bibinfo  {journal} {Journal of High
  Energy Physics}\ }\textbf {\bibinfo {volume} {2017}},\ \bibinfo {pages} {48}
  (\bibinfo {year} {2017})}\BibitemShut {NoStop}%
\bibitem [{\citenamefont {Kos}\ \emph {et~al.}(2018)\citenamefont {Kos},
  \citenamefont {Ljubotina},\ and\ \citenamefont {Prosen}}]{Prosen2018RMT}%
  \BibitemOpen
  \bibfield  {author} {\bibinfo {author} {\bibfnamefont {P.}~\bibnamefont
  {Kos}}, \bibinfo {author} {\bibfnamefont {M.}~\bibnamefont {Ljubotina}}, \
  and\ \bibinfo {author} {\bibfnamefont {T.}~\bibnamefont {Prosen}},\ }\href
  {\doibase 10.1103/PhysRevX.8.021062} {\bibfield  {journal} {\bibinfo
  {journal} {Physical Review X}\ }\textbf {\bibinfo {volume} {8}},\ \bibinfo
  {pages} {021062} (\bibinfo {year} {2018})}\BibitemShut {NoStop}%
\bibitem [{\citenamefont {{Benjamin}}\ \emph {et~al.}(2019)\citenamefont
  {{Benjamin}}, \citenamefont {{Dyer}}, \citenamefont {{Fitzpatrick}},\ and\
  \citenamefont {{Xin}}}]{Benjamin2019irrational}%
  \BibitemOpen
  \bibfield  {author} {\bibinfo {author} {\bibfnamefont {N.}~\bibnamefont
  {{Benjamin}}}, \bibinfo {author} {\bibfnamefont {E.}~\bibnamefont {{Dyer}}},
  \bibinfo {author} {\bibfnamefont {A.~L.}\ \bibnamefont {{Fitzpatrick}}}, \
  and\ \bibinfo {author} {\bibfnamefont {Y.}~\bibnamefont {{Xin}}},\ }\href
  {\doibase 10.1007/JHEP04(2019)025} {\bibfield  {journal} {\bibinfo  {journal}
  {Journal of High Energy Physics}\ }\textbf {\bibinfo {volume} {2019}},\
  \bibinfo {eid} {25} (\bibinfo {year} {2019})}\BibitemShut {NoStop}%
\bibitem [{\citenamefont {Bohigas}\ and\ \citenamefont
  {Giannoni}(1984)}]{Bohigas1984}%
  \BibitemOpen
  \bibfield  {author} {\bibinfo {author} {\bibfnamefont {O.}~\bibnamefont
  {Bohigas}}\ and\ \bibinfo {author} {\bibfnamefont {M.-J.}\ \bibnamefont
  {Giannoni}},\ }in\ \href@noop {} {\emph {\bibinfo {booktitle} {Mathematical
  and Computational Methods in Nuclear Physics}}},\ \bibinfo {editor} {edited
  by\ \bibinfo {editor} {\bibfnamefont {J.~S.}\ \bibnamefont {Dehesa}},
  \bibinfo {editor} {\bibfnamefont {J.~M.~G.}\ \bibnamefont {Gomez}}, \ and\
  \bibinfo {editor} {\bibfnamefont {A.}~\bibnamefont {Polls}}}\ (\bibinfo
  {publisher} {Springer Berlin Heidelberg},\ \bibinfo {address} {Berlin,
  Heidelberg},\ \bibinfo {year} {1984})\ pp.\ \bibinfo {pages}
  {1--99}\BibitemShut {NoStop}%
\bibitem [{\citenamefont {Turner}\ \emph {et~al.}(2018)\citenamefont {Turner},
  \citenamefont {Michailidis}, \citenamefont {Abanin}, \citenamefont {Serbyn},\
  and\ \citenamefont {Papi{\'c}}}]{Turner2018scar}%
  \BibitemOpen
  \bibfield  {author} {\bibinfo {author} {\bibfnamefont {C.~J.}\ \bibnamefont
  {Turner}}, \bibinfo {author} {\bibfnamefont {A.~A.}\ \bibnamefont
  {Michailidis}}, \bibinfo {author} {\bibfnamefont {D.~A.}\ \bibnamefont
  {Abanin}}, \bibinfo {author} {\bibfnamefont {M.}~\bibnamefont {Serbyn}}, \
  and\ \bibinfo {author} {\bibfnamefont {Z.}~\bibnamefont {Papi{\'c}}},\ }\href
  {\doibase 10.1038/s41567-018-0137-5} {\bibfield  {journal} {\bibinfo
  {journal} {Nature Physics}\ }\textbf {\bibinfo {volume} {14}},\ \bibinfo
  {pages} {745} (\bibinfo {year} {2018})}\BibitemShut {NoStop}%
\bibitem [{\citenamefont {Moudgalya}\ \emph {et~al.}(2018)\citenamefont
  {Moudgalya}, \citenamefont {Regnault},\ and\ \citenamefont
  {Bernevig}}]{Moudgalya2018scar}%
  \BibitemOpen
  \bibfield  {author} {\bibinfo {author} {\bibfnamefont {S.}~\bibnamefont
  {Moudgalya}}, \bibinfo {author} {\bibfnamefont {N.}~\bibnamefont {Regnault}},
  \ and\ \bibinfo {author} {\bibfnamefont {B.~A.}\ \bibnamefont {Bernevig}},\
  }\href {\doibase 10.1103/PhysRevB.98.235156} {\bibfield  {journal} {\bibinfo
  {journal} {Physical Review B}\ }\textbf {\bibinfo {volume} {98}},\ \bibinfo
  {pages} {235156} (\bibinfo {year} {2018})}\BibitemShut {NoStop}%
\bibitem [{\citenamefont {Lin}\ and\ \citenamefont
  {Motrunich}(2019)}]{Lin2019scar}%
  \BibitemOpen
  \bibfield  {author} {\bibinfo {author} {\bibfnamefont {C.-J.}\ \bibnamefont
  {Lin}}\ and\ \bibinfo {author} {\bibfnamefont {O.~I.}\ \bibnamefont
  {Motrunich}},\ }\href {\doibase 10.1103/PhysRevLett.122.173401} {\bibfield
  {journal} {\bibinfo  {journal} {Physical Review Letters}\ }\textbf {\bibinfo
  {volume} {122}},\ \bibinfo {pages} {173401} (\bibinfo {year}
  {2019})}\BibitemShut {NoStop}%
\bibitem [{\citenamefont {Hannay}\ and\ \citenamefont
  {Berry}(1980)}]{Hannay-Berry}%
  \BibitemOpen
  \bibfield  {author} {\bibinfo {author} {\bibfnamefont {J.}~\bibnamefont
  {Hannay}}\ and\ \bibinfo {author} {\bibfnamefont {M.}~\bibnamefont {Berry}},\
  }\href {\doibase https://doi.org/10.1016/0167-2789(80)90026-3} {\bibfield
  {journal} {\bibinfo  {journal} {Physica D: Nonlinear Phenomena}\ }\textbf
  {\bibinfo {volume} {1}},\ \bibinfo {pages} {267 } (\bibinfo {year}
  {1980})}\BibitemShut {NoStop}%
\bibitem [{\citenamefont {Chertkov}\ \emph {et~al.}(2020)\citenamefont
  {Chertkov}, \citenamefont {Villalonga},\ and\ \citenamefont
  {Clark}}]{Chertkov2020appendixH}%
  \BibitemOpen
  \bibfield  {author} {\bibinfo {author} {\bibfnamefont {E.}~\bibnamefont
  {Chertkov}}, \bibinfo {author} {\bibfnamefont {B.}~\bibnamefont
  {Villalonga}}, \ and\ \bibinfo {author} {\bibfnamefont {B.~K.}\ \bibnamefont
  {Clark}},\ }\href {\doibase 10.1103/PhysRevResearch.2.023348} {\bibfield
  {journal} {\bibinfo  {journal} {Phys. Rev. Research}\ }\textbf {\bibinfo
  {volume} {2}},\ \bibinfo {pages} {023348} (\bibinfo {year}
  {2020})}\BibitemShut {NoStop}%
\bibitem [{\citenamefont {Keating}(1991)}]{Keating1991}%
  \BibitemOpen
  \bibfield  {author} {\bibinfo {author} {\bibfnamefont {J.~P.}\ \bibnamefont
  {Keating}},\ }\href {\doibase 10.1088/0951-7715/4/2/006} {\bibfield
  {journal} {\bibinfo  {journal} {Nonlinearity}\ }\textbf {\bibinfo {volume}
  {4}},\ \bibinfo {pages} {309} (\bibinfo {year} {1991})}\BibitemShut {NoStop}%
\bibitem [{\citenamefont {Esposti}\ and\ \citenamefont
  {Winn}(2005)}]{Esposti-Winn}%
  \BibitemOpen
  \bibfield  {author} {\bibinfo {author} {\bibfnamefont {M.~D.}\ \bibnamefont
  {Esposti}}\ and\ \bibinfo {author} {\bibfnamefont {B.}~\bibnamefont {Winn}},\
  }\href {\doibase 10.1088/0305-4470/38/26/005} {\bibfield  {journal} {\bibinfo
   {journal} {Journal of Physics A: Mathematical and General}\ }\textbf
  {\bibinfo {volume} {38}},\ \bibinfo {pages} {5895} (\bibinfo {year}
  {2005})}\BibitemShut {NoStop}%
\bibitem [{\citenamefont {Degli~Esposti}\ and\ \citenamefont
  {Graffi}(2003)}]{Backer2003book}%
  \BibitemOpen
  \bibfield  {author} {\bibinfo {author} {\bibfnamefont {M.}~\bibnamefont
  {Degli~Esposti}}\ and\ \bibinfo {author} {\bibfnamefont {S.}~\bibnamefont
  {Graffi}},\ }in\ \href@noop {} {\emph {\bibinfo {booktitle} {The mathematical
  aspects of quantum maps}}}\ (\bibinfo  {publisher} {Springer},\ \bibinfo
  {year} {2003})\ pp.\ \bibinfo {pages} {49--90}\BibitemShut {NoStop}%
\bibitem [{\citenamefont {Waters}(1982)}]{WaltersTextbook}%
  \BibitemOpen
  \bibfield  {author} {\bibinfo {author} {\bibfnamefont {P.}~\bibnamefont
  {Waters}},\ }\href@noop {} {\emph {\bibinfo {title} {An Introduction to
  Ergodic Theory}}}\ (\bibinfo  {publisher} {Springer New York},\ \bibinfo
  {year} {1982})\BibitemShut {NoStop}%
\bibitem [{Note1()}]{Note1}%
  \BibitemOpen
  \bibinfo {note} {We perform the unfolding procedure to the energy spectrum
  such that the mean level spacing is one everywhere. See~\cite
  {Guhr1998unfolding} for details.}\BibitemShut {Stop}%
\bibitem [{Note2()}]{Note2}%
  \BibitemOpen
  \bibinfo {note} {Due to the complication of $A_{12} \not =1$, we are not able
  to simplify the time evolution operator~\protect \textup {\hbox
  {\mathsurround \z@ \protect \normalfont (\ignorespaces \ref {eq:
  perturbedCatU14131514}\unskip \@@italiccorr )}} as was done in~\cite
  {Shivaji2019}, therefore are limited to relative small system size when
  evaluating coefficient entropy.}\BibitemShut {Stop}%
\bibitem [{\citenamefont {github.com/ClarkResearchGroup/qosy}()}]{qosy}%
  \BibitemOpen
  \bibfield  {author} {\bibinfo {author} {\bibnamefont
  {github.com/ClarkResearchGroup/qosy}},\ }\href@noop {} {\ }\BibitemShut
  {NoStop}%
\bibitem [{\citenamefont {Ireland}\ and\ \citenamefont
  {Rosen}(1990)}]{IrelandBook}%
  \BibitemOpen
  \bibfield  {author} {\bibinfo {author} {\bibfnamefont {K.}~\bibnamefont
  {Ireland}}\ and\ \bibinfo {author} {\bibfnamefont {M.}~\bibnamefont
  {Rosen}},\ }\href@noop {} {\emph {\bibinfo {title} {A Classical Introduction
  to Modern Number Theory}}}\ (\bibinfo  {publisher} {Springer New York},\
  \bibinfo {year} {1990})\BibitemShut {NoStop}%
\bibitem [{\citenamefont {Guhr}\ \emph {et~al.}(1998)\citenamefont {Guhr},
  \citenamefont {M{\"u}ller-Groeling},\ and\ \citenamefont
  {Weidenm{\"u}ller}}]{Guhr1998unfolding}%
  \BibitemOpen
  \bibfield  {author} {\bibinfo {author} {\bibfnamefont {T.}~\bibnamefont
  {Guhr}}, \bibinfo {author} {\bibfnamefont {A.}~\bibnamefont
  {M{\"u}ller-Groeling}}, \ and\ \bibinfo {author} {\bibfnamefont {H.~A.}\
  \bibnamefont {Weidenm{\"u}ller}},\ }\href {\doibase
  https://doi.org/10.1016/S0370-1573(97)00088-4} {\bibfield  {journal}
  {\bibinfo  {journal} {Physics Reports}\ }\textbf {\bibinfo {volume} {299}},\
  \bibinfo {pages} {189} (\bibinfo {year} {1998})}\BibitemShut {NoStop}%
\end{thebibliography}%

\end{document}